\documentclass[12pt]{article}

\DeclareUnicodeCharacter{2009}{\,}
\DeclareUnicodeCharacter{202F}{\,}
\DeclareUnicodeCharacter{2212}{-}  

\usepackage[utf8]{inputenc}
\usepackage{graphicx}
\usepackage{amsmath, amssymb}
\usepackage{geometry}
\usepackage{caption}
\usepackage{subcaption}
\usepackage{booktabs}
\usepackage{multirow}
\usepackage{textgreek}
\usepackage{mhchem}
\usepackage{float}
\usepackage{siunitx}
\usepackage{threeparttable}

\usepackage[numbers,sort&compress]{natbib}
\setcitestyle{numbers,square}

\usepackage{hyperref}

\begin{document}

\title{ Stability Analysis and Optoelectronic Properties of Mg$_3$ZBr$_3$ (Z = As, Sb, Bi) Perovskites for Evaluating the Performance in PIN Photo Diode} 

\author{
Md Mohiuddin\textsuperscript{1,$\Upsilon$}, Mohammed Mehedi Hasan\textsuperscript{2,$\Upsilon$}, Alamgir Kabir\textsuperscript{1,*}}

\date{}

\maketitle

\begin{center}
    \textsuperscript{1}Department of Physics, University of Dhaka, Dhaka 1000, Bangladesh. \\
    \textsuperscript{2}Department of Physics \& Astronomy, Louisiana State University, Baton Rouge, LA 70803, USA. \\
    \textsuperscript{*}Corresponding author: alamgir.kabir@du.ac.bd \\
    \textsuperscript{$\Upsilon$}Equal Contributions
\end{center}

\begin{abstract}
  The toxicity and stability issues of lead-based perovskites motivate non-toxic, durable alternatives. This work examines lead-free Mg$_3$ZBr$_3$ (Z = As, Sb, Bi) halide perovskites as optoelectronic materials, with emphasis on Mg$_3$AsBr$_3$ and Mg$_3$SbBr$_3$. All Mg$_3$ZBr$_3$ perovskites are predicted from first-principles calculations to crystallise in the cubic $Pm\bar{3}m$ phase, indirect gaps of $2.0645$\,eV for Mg$_3$AsBr$_3$, $1.6533$\,eV for Mg$_3$SbBr$_3$ and $1.5226 $\,eV for Mg$_3$BiBr$_3$ were observed by hybrid functional. Optical spectra show a rise in absorption above the gap and an increasing static dielectric response along As$\!\to\!$Sb$\!\to\!$Bi. Phonon dispersion lacks imaginary branches for Mg$_3$AsBr$_3$ and Mg$_3$SbBr$_3$ , indicates the dynamical stability of these two materials under consideration, and exhibits large mode anharmonicity (Grüneisen signatures), consistent with soft-lattice heat transport trends. It has been found, moving down the pnictogen series expands the lattice and lowers the Goldschmidt tolerance factor, which, together with enhanced pnictogen–Br $p$-orbital hybridization and stereochemically active $ns^2$ lone pairs (Sb, Bi), narrows the band gap and elevates the optical dielectric response. Elastic analyses confirm Born stability and moderate stiffness, with Hill-averaged bulk moduli decreasing from $\sim\!44$\,GPa (Mg$_3$AsBr$_3$) to $\sim\!35$\,GPa (Mg$_3$BiBr$_3$). Drift–diffusion \emph{p}–\emph{i}–\emph{n} simulations qualitatively track band-edge–limited spectra, aligning with the computed gaps. Together, these results position these materials as lead-free candidates for stable thin-film photodiodes and photovoltaics applications.


\end{abstract}

\section*{Introduction}
Metal-halide perovskites have proven to be one of the most transformative material families in modern optoelectronics and are revolutionising photovoltaic (PV) technologies. The power conversion efficiencies (PCEs) of perovskite solar cell (PSC) have reached over 27\% in 2025 \cite{diercks2025sequential}. Previously, it was first reported to be of 3.8\% in 2009, competing with crystalline silicon while offering advantages in solution processability, tunable bandgaps, and low-temperature fabrication \cite{Kojima,Jin}. Though perovskites have been studied a lot both theoretically and experimentally because of high absorption coefficients ($>10^4 \, \text{cm}^{-1}$), ambipolar charge transport, long carrier diffusion lengths ($>1 \, \mu\text{m}$), and high defect tolerance but it has been a point of major concern to find functional perovskites which is non-toxic to the environment \cite{Green2014, Stranks2013,Zhang2018}. Additionally, instability under moisture, heat, and light limits operational lifetimes, necessitating encapsulation strategies that increase costs \cite{Zhang2022,Liu2025,Leijtens2013}. To address these issues, lead-free halide perovskites have drawn attention because of their stability, excellent performance in devices, and eco-friendly attributes. Investigations into these materials are underway for a variety of applications that extend beyond solar cells, such as energy harvesting, radiation detection, light-emitting diodes (LEDs), photodetectors, displays, photocatalysis, resistive memory devices, and sensors \cite{mukhtar2025eco, afroz2023design, aftab2023quantum}.

One of the emerging classes of lead-free perovskites is the $A_3ZX_3$ family, where $A$ and $Z$ represent large and small  cations ($\text{Mg}^{2+}$, $\text{Ba}^{2+}$, $\text{Sr}^{2+}$, $\text{Ca}^{2+}$, $\text{Sb}^{3+}$, $\text{As}^{3+}$, $\text{P}^{3+}$), respectively, and $X$ represents $\text{F}^-$, $\text{Cl}^-$, $\text{Br}^-$, or $\text{I}^-$ which allow compositional flexibility to tailor electronic and structural properties \cite{mayeen2025,Snaith2013}. In an extensive study, Feng et al. reported the bandgaps of the $A_3MX_3$ (A = Mg, Ca, Sr, Ba; M = N, P, As, Sb; X = F, Cl, Br, I) materials for deciding the efficient materials in photovoltaics based on the theoretical efficiency and thermal stability \cite{feng2021}.

Bismuth ($\text{Bi}^{3+}$)-based double perovskites (e.g., $\text{Cs}_2\text{AgBiBr}_6$) offer better stability but suffer from indirect bandgaps and low carrier mobility, limiting PCEs to $<5\%$ as reported by Slavney et al. \cite{Slavney2016}. Zhang and co-authors have shown that stability can be improved with defect engineering and dimensional reduction (2D/3D heterostructures) , yet efficiencies remain quite low compared to lead-based counterparts \cite{Zhang2021}.On the other hand, Tin ($\text{Sn}^{2+}$)-halide perovskites, such as $\text{CsSnI}_3$, initially showed promise with bandgaps near 1.3 eV and PCEs up to 11.57\% \cite{Liu2023}. However, $\text{Sn}^{2+}$ oxidizes rapidly to $\text{Sn}^{4+}$, creating vacancies that degrade performance \cite{marshall2015}. The band gap of perovskites is dependant on the the $BX_6$ octahedral framework\cite{Lee2016}, it is shown that in $\text{MAPbI}_3$, the Pb-I-Pb bond angles (165°--180°) and octahedral tilting dictate carrier mobility and bandgap \cite{Filip2014,Motta2015}. From the reports of Zhang et al it can be observed that substituting $\text{Pb}^{2+}$ with smaller $\text{Mg}^{2+}$ induces lattice contraction, altering tilting dynamics\cite{Zhang2018_tilting}. Experimental X-ray diffraction confirms this lattice contraction—reflected by a 0.05–0.07 degree shift of the (110) and (220) reflections—while preserving the perovskite phase validating the theoretical expectations\cite{yang2018}. Furthermore, 24.73\% PEC has been observed for $Ca_3NCl_3$ as absorber which demonstrated a good response to visible spectrum absorption\cite{RAHMAN2025131873}. In another study, Apurba et al. demonstrated the direct bandgap and stability of $Mg_3SbX_3$(X = I, Br, Cl, F) as a promising material for photovoltaic applications\cite{Apurba2024}. Magnesium ($\text{Mg}^{2+}$)-based halide perovskites represent an relatevely underexplored but promising avenue. With an ionic radius comparable to $\text{Pb}^{2+}$, $\text{Mg}^{2+}$ can occupy octahedral sites while offering non-toxicity and earth abundance \cite{Shannon1976}. Early studies on $\text{CsMgX}_3$ ($X = \text{Cl}$, $\text{Br}$) revealed indirect bandgaps of 6.35 and 4.26 eV, respectively, which is suitable for UV photodetectors and dielectric applications \cite{Kaewmeechai_2019,SHARMA2021113415}. In addition to that Sharma et al. predicted that $\text{CsMgBr}_3$ is an indirect bandgap material with moderate thermoelectric performance ($ZT > 0.79$) \cite{SHARMA2021113415}.

This study addresses two critical gaps. By integrating As, Sb, and Bi into the material design, toxicity is avoided, aligning with global compliance standards. The second reason is the high lattice energy of magnesium and the strong $\text{Mg-Br}$ bonds that can enhance structural stability under thermal or mechanical stress \cite{Darwent1970,parosh2025, Sotudesh2022}. Using density functional theory (DFT), we systematically analyze $\text{Mg}_3\text{ZBr}_3$ ($Z = \text{As}$, $\text{Sb}$, $\text{Bi}$) perovskites with the aim of elucidating structural, electronic, optical, and mechanical properties. Furthermore, we studied dynamical stability and assessed their viability as lead-free perovskites via PIN junction photodiode device simulations. This study provides the first unified analysis of $\text{Mg}_3\text{ZBr}_3$ perovskites, bridging structural and optoelectronic properties and making them potential candidates in optoelectronic applications.

\section{Computational Methodology}

All first-principles calculations were performed using the Vienna \emph{Ab Initio} Simulation Package (VASP) v6.1.0 \cite{Kresse1996}. VASP input files were generated and analyzed with the Python Materials Genomics library (pymatgen) \cite{Ong2013} and the Atomic Simulation Environment (ASE) \cite{Larsen2017}. We employed the projector-augmented wave (PAW) method \cite{Blochl1994} with VASP’s standard PBE-PAW datasets. The valence electrons treated for each species were Mg $2p^6 3s^2$, As $4s^2 4p^3$, Sb $5s^2 5p^3$, Bi $6s^2 6p^3$, and Br $4p^5$. Exchange–correlation effects were described by the Perdew–Burke–Ernzerhof (PBE) generalized gradient approximation (GGA) \cite{Perdew1996} for geometry optimizations and most property calculations, while the screened hybrid functional of Heyd, Scuseria, and Ernzerhof (HSE06) \cite{Krukau2006} was used for high-accuracy structural relaxation and band structure computations. A plane-wave kinetic energy cutoff of 300~eV was used, exceeding the recommended minimum for all involved PAW potentials. Brillouin-zone integrations employed Monkhorst–Pack $k$-point meshes \cite{Monkhorst1976} of $6\times6\times6$ for the primitive cubic Mg$_3$ZBr$_3$ unit cells, with commensurate mesh densities for larger supercells. All structures were fully relaxed with respect to lattice parameters and atomic positions using the conjugate-gradient algorithm until residual Hellmann–Feynman forces fell below 0.01eV Å$^{-1}$ on each atom.
 High accuracy band structures were obtained by non-self-consistent evaluations using the HSE06 functional (25\% exact exchange with a screening parameter of 0.2~Å$^{-1}$ \cite{Krukau2006}). Elastic properties were derived from stress–strain calculations at the PBE level, by applying small deformations to the lattice and extracting the elastic stiffness tensor from the resulting stress responses. Optical properties, specifically the frequency-dependent dielectric function and absorption spectrum, were calculated within the independent-particle approximation using the PBE functional. Lattice-dynamical properties were evaluated using density-functional perturbation theory (DFPT) \cite{Baroni2001} as implemented in VASP. Phonon dispersion relations and phonon densities of states were then obtained by post-processing the interatomic force constants with the Phonopy code \cite{Togo2015}.

Device simulations were performed in COMSOL Multiphysics (Semiconductor and EWFD Module) using a stationary drift--diffusion formulation under DC reverse bias. The device was treated as one-dimensional with domain length $h_{\mathrm{dom}}=1~\mu$m. Ohmic p and n contacts were applied at the two ends. Gaussian dopant profiles were used at each contact with peak concentrations $N_A^{\max}=N_D^{\max}=1\times10^{20}~\mathrm{cm^{-3}}$ and junction depth $0.1\,h_{\mathrm{dom}}=0.1~\mu$m; the interior was intrinsic.

\section{Structural Properties}
\begin{figure}
    \centering
    \includegraphics[width=1\linewidth]{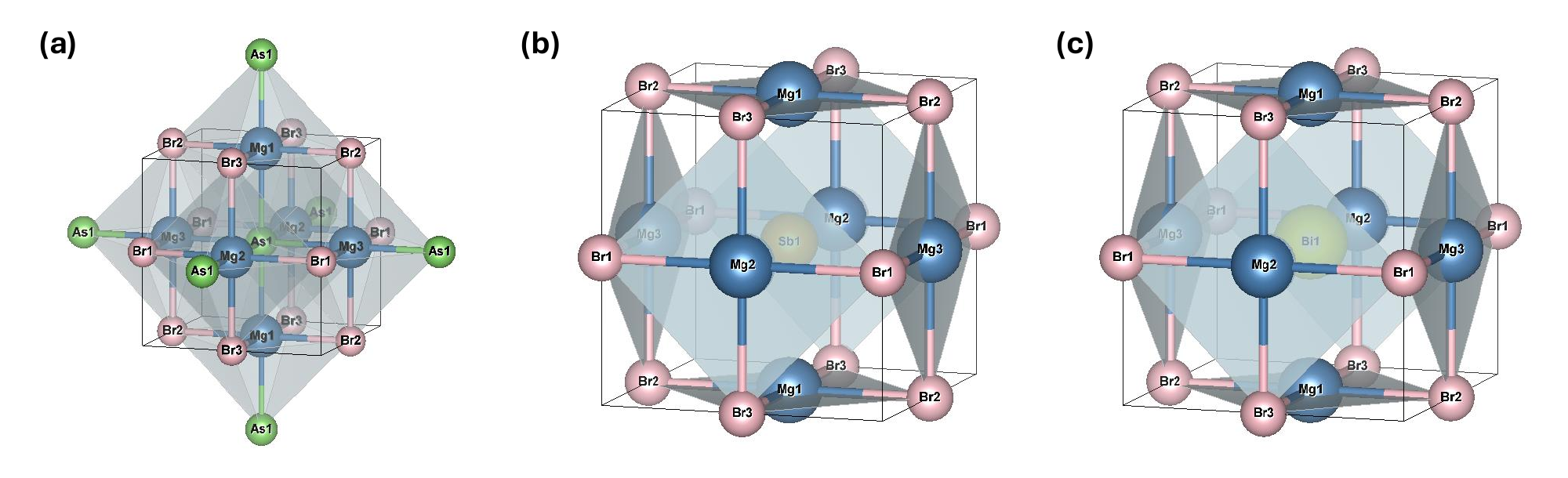}
    \caption{Crystal structures of (a) Mg$_{3}$AsBr$_{3}$ with As atoms (green) surrounded by Br atoms (pink) with Mg atoms (blue) at the corners, (b) Mg$_{3}$SbBr$_{3}$ with Bi atom (yellow) at the center of the octahedron and (c) Mg$_{3}$BiBr$_{3}$ with Sb atom (yellow) at the center of the octahedron.}
    \label{fig:crystals}
    
\end{figure}

\begin{figure}
    \centering
    \includegraphics[width=1.0\linewidth]{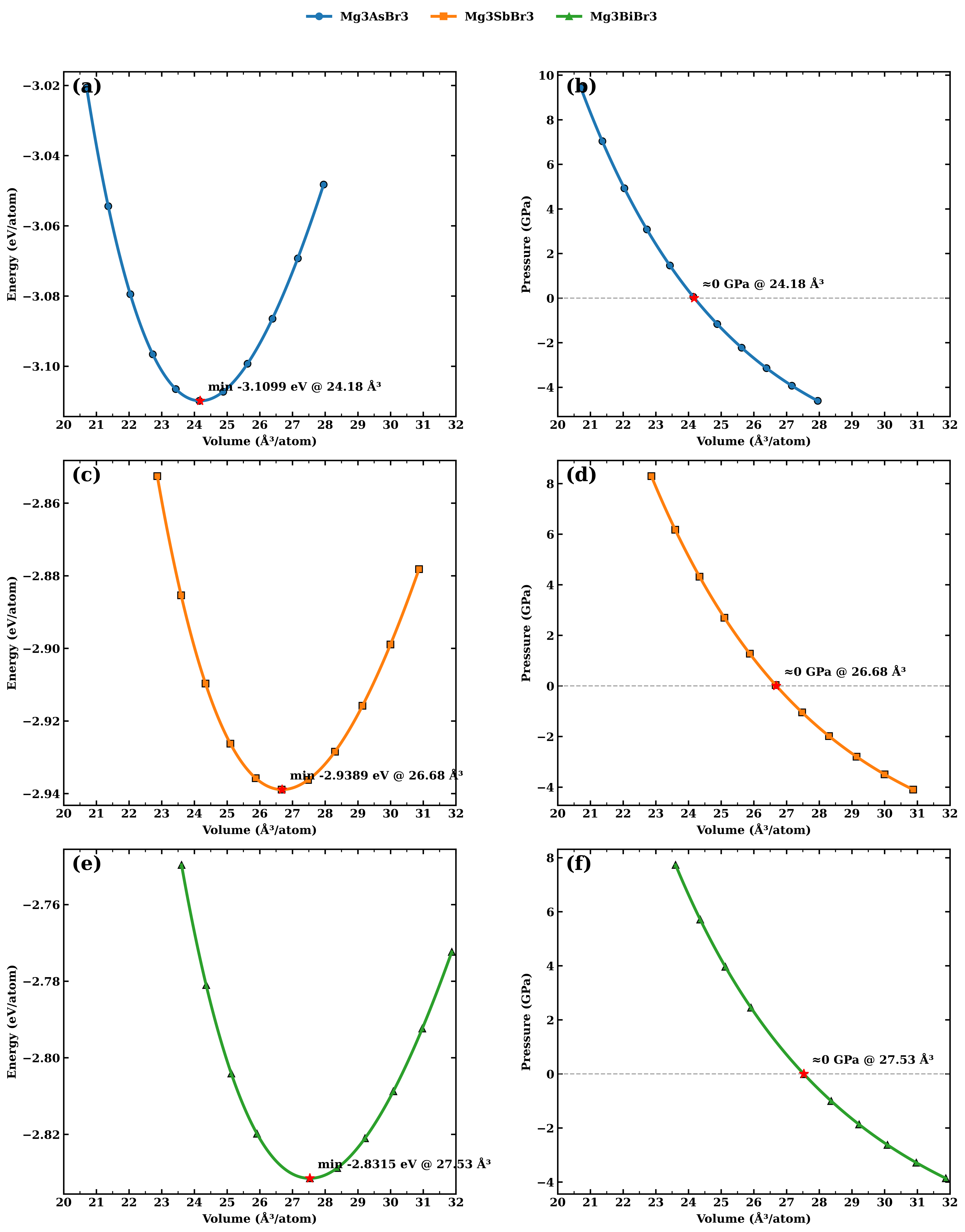}
    \caption{Equation-of-state (EOS) analysis for the Mg\(_3\)AsBr\(_3\),
    Mg\(_3\)SbBr\(_3\), and Mg\(_3\)BiBr\(_3\) perovskites.
    Here (a),(c) and (e) correspond to the  total energy per atom versus volume, respectively.  The red star indicates the energy minimum. Here (b),(d) and (e) subplot plots display the corresponding pressure–volume curves; the red star indicates the zero-pressure crossing that coincides with \(V_0\). Dashed grey lines denote the zero-pressure reference.}
    \label{fig:EOS_Mg3XBr3}
    
\end{figure}

The halide perovskites Mg$_3$AsBr$_3$, Mg$_3$SbBr$_3$, and Mg$_3$BiBr$_3$ crystallise in the simple‐cubic space group $Pm\bar{3}m$ (\#221), as show in Figure~\ref{fig:crystals}.  Within this topology the pnictogen centre occupies the body–centre site $(\tfrac12,\tfrac12,\tfrac12)$, magnesium resides on the three face‐centred positions $(\tfrac12,\tfrac12,0)$, $(0,\tfrac12,\tfrac12)$, and $(\tfrac12,0,\tfrac12)$, while bromine terminates the cube at $(0,0,\tfrac12)$, $(\tfrac12,0,0)$, and $(0,\tfrac12,0)$, the optimized lattice parameters are reported in Table-\ref{tab:materials_data}. It is evident from Table~\ref{tab:materials_data} that both the lattice parameter and unit-cell volume increase progressively from Mg$_3$AsBr$_3$ to Mg$_3$SbBr$_3$ and then to Mg$_3$BiBr$_3$.This monotonic expansion reflects the progressive increase in the pnictogen ionic radius down group~15 of the periodic table \citep{Shannon1976}. Complementary equation-of-state calculations, summarised in Figure~\ref{fig:EOS_Mg3XBr3}, show equilibrium volumes $V_0$ of \SI{24.18}{\angstrom^{3},atom^{-1}} for Mg$_3$AsBr$_3$ (Figure~2-a), \SI{26.68}{\angstrom^{3},atom^{-1}} for Mg$_3$SbBr$_3$ (Figure~2-b), and \SI{27.53}{\angstrom^{3},atom^{-1}} for Mg$_3$BiBr$_3$ (Figure~2-c). Concurrently, the binding energy per atom at equilibrium becomes less negative: Mg$_3$AsBr$_3$ is the most strongly bound of the three, while Mg$_3$BiBr$_3$ exhibits the weakest binding. These observations indicate that substituting larger pnictogen atoms results in softer lattices and potentially lower bulk moduli, as further confirmed in Table~\ref{tab:merged_mechanical_swapped}.

\begin{table}[h!]
    \centering
    \caption{Optimised cell parameters ($a$) and volumes ($V$) for Mg$_3$ZBr$_3$ (Z~=~As, Sb, Bi) obtained with the PBE and screened–hybrid HSE06 functionals. The tolerance factor $t$ is listed for each Z.}
    \label{tab:materials_data}
    \begin{tabular}{@{}lllll@{}}
        \toprule
        \textbf{Compound} & \textbf{$a$ (\AA)} & \textbf{$V$ (\AA$^{3}$)} & \textbf{Method} & \textbf{Tolerance Factor } \\
        \midrule
        \multirow{2}{*}{Mg$_3$AsBr$_3$} & 5.529 & 169.000 & PBE   & \multirow{2}{*}{0.78} \\
                                        & 5.480 & 164.566 & HSE06 &                        \\
        \cmidrule{1-5}
        \multirow{2}{*}{Mg$_3$SbBr$_3$} & 5.715 & 186.679 & PBE   & \multirow{2}{*}{0.74} \\
                                        & 5.660 & 181.321 & HSE06 &                        \\
        \cmidrule{1-5}
        \multirow{2}{*}{Mg$_3$BiBr$_3$} & 5.776 & 192.732 & PBE   & \multirow{2}{*}{0.63} \\
                                        & 5.718 & 186.953 & HSE06 &                        \\
        \bottomrule
    \end{tabular}
\end{table}

Although the unit‐cell edge length grows steadily, the incremental volume increase between Sb and Bi members (\SI{186.7}{\angstrom^{3}}\,$\rightarrow$\,\SI{192.7}{\angstrom^{3}}) is smaller than naively expected from ionic‐radius arguments alone.  Subtle changes in metal–halogen covalency and electronic configuration appear to temper the volumetric response, an observation echoed in earlier perovskite surveys\citep{BURDETT1984173,Walsh2015}. Electronic‐structure insights follow logically from these metrics.  A wider gap is anticipated for Mg$_3$AsBr$_3$, driven by the compact As$^{5+}$\,4$p$ manifold, whereas the Bi‐rich phase should exhibit improved carrier transport via spatially extended 6$p$ orbitals\citep{Walsh2015}.  Lone‐pair activity on Sb$^{3+}$ and Bi$^{3+}$ may also induce second‐order Jahn–Teller displacements that subtly tailor the band topology\citep{Walsh2011}.  Such features, coupled with the benign defect chemistry familiar in halide perovskites\citep{Brandt2015}, argue for continued exploration of these Mg–based materials as environmentally friendly absorbers in photovoltaic and optoelectronic devices\citep{Green2014}.

Goldschmidt tolerance factors ($t$) provide a rapid gauge of perovskite formability.  Employing Shannon’s six‐coordination radii ($r_{\mathrm{Mg}^{2+}}$~=~\SI{0.72}{\angstrom}; $r_{\mathrm{Br}^{-}}$~=~\SI{1.96}{\angstrom}; $r_{\mathrm{As}^{5+}}$~=~\SI{0.46}{\angstrom}; $r_{\mathrm{Sb}^{5+}}$~=~\SI{0.60}{\angstrom}; $r_{\mathrm{Bi}^{3+}}$~=~\SI{1.03}{\angstrom}) gives $t$ values of 0.78, 0.74, and 0.63 for Z~=~As, Sb, and Bi, respectively.  All three lie below the canonical cubic interval (0.90–1.00)\citep{Bartel2019}, forecasting octahedral tilts or lower‐symmetry phases at ambient conditions\citep{Radha2018}.  The associated octahedral parameters $\mu=r_B/r_X$ equal 0.23, 0.31, and 0.53, indicating that Bi substitution should deliver the most mechanically robust framework, whereas the As analogue approaches the threshold of geometric instability\citep{Travis2016}. The tolerance‐factor analysis signals latent structural flexibility that could be harnessed for phase‐change sensing or chromic platforms.  
\section{Phonon and Lattice Dynamics}

\begin{figure}[tbhp]
\centering
\includegraphics[width=1.0\textwidth]{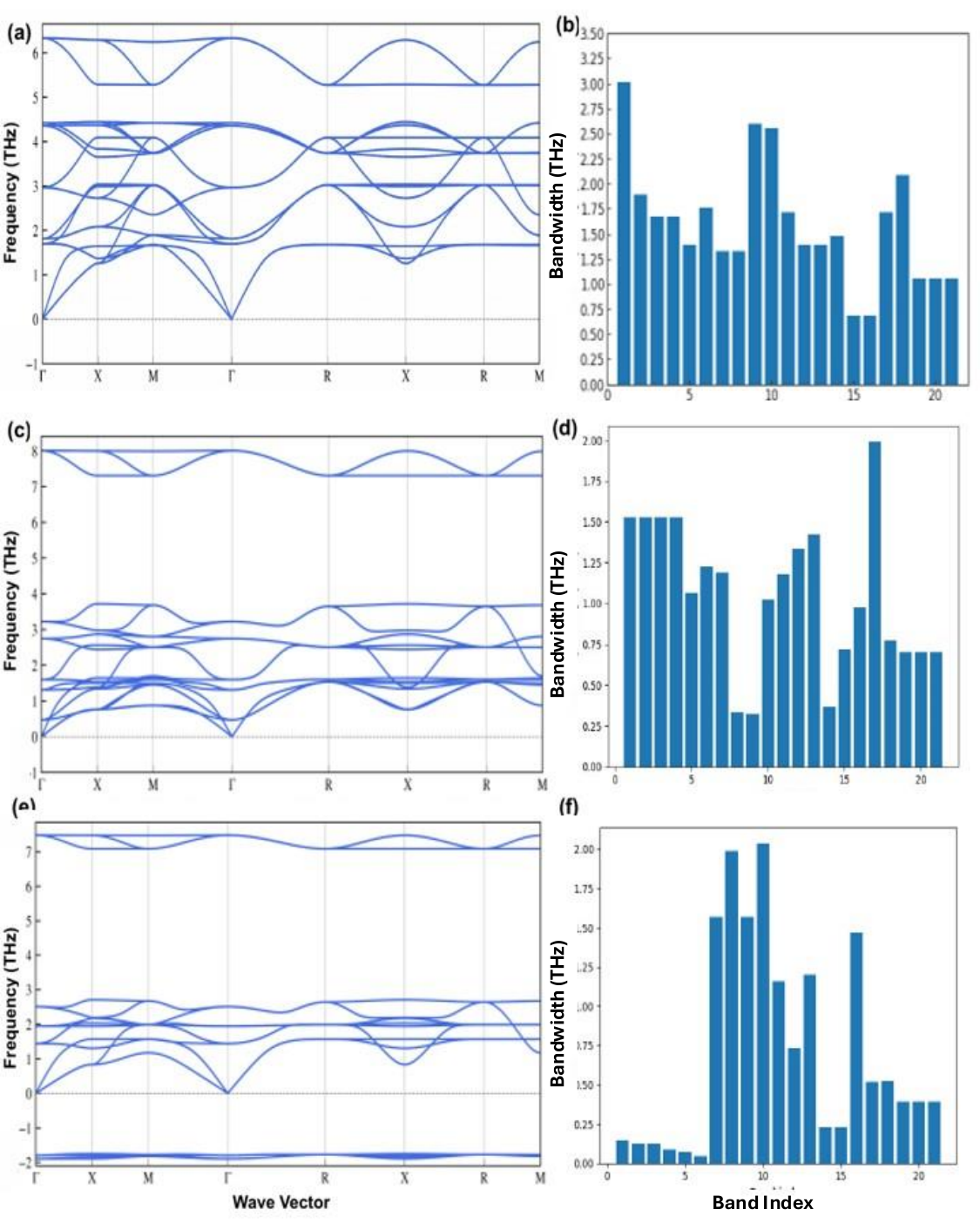}
\caption{Phonon dispersion relations of cubic Mg$_3Z$Br$_3$ (Z = As, Sb, Bi) along high-symmetry directions and the frequency range that each branch spans along the  computed path (a) - (b) for  Mg$_3$AsBr$_3$; (c) - (d) Mg$_3$SbBr$_3$; (e) - (f) Mg$_3$BiBr$_3$.  }
\label{fig:phonon_bands}
\end{figure}

The phonon dispersion curves of cubic Mg$_3Z$Br$_3$ (Z = As, Sb, Bi) are shown in Figure~\ref{fig:phonon_bands}. It is clear from Figure~\ref{fig:phonon_bands} (a and c) that Mg$_3As$Br$_3$ and Mg$_3Sb$Br$_3$ are dynamically stable in the cubic phase at 0K, as evidenced by the absence of any imaginary  phonon modes in the calculated dispersions \cite{Nepal2024-Y2C3-phonons}.But the presence of imaginary phonon bands for Mg$_3Bi$Br$_3$ indicates that the structure is dynamically unstable.  

For Z = As, Sb, the phonon bandwidth (highest optical frequency) increases gradually from about $6.33$~THz to $8.00$ THz (Figure~\ref{fig:phonon_bands} a and c). This trend is expected because heavier atoms and larger unit-cell volumes generally lead to lower vibrational frequencies\cite{Khandaker2025-Mg3AB3}. In fact, a clear decline in Debye temperature $\theta_D$ (which scales with the acoustic phonon frequencies) is observed as the atomic mass increases from As to Sb to Bi, which can be verified from the Hill-averaged mechanical properties in table \ref{tab:merged_mechanical_swapped}. The acoustic branches near the Brillouin-zone center ($\Gamma$) become progressively less steep from As to Bi, indicating a reduction in sound velocities for the heavier compounds. This is consistent with the heavier average atomic mass and expanded lattice of Mg$_3$BiBr$_3$, which results in lower elastic wave speeds\cite{Haeger2020-halide-thermal}. The phonon bandwidth analysis of Figure~\ref{fig:phonon_bands} (b, d and f) reveals a systematic reduction in the frequency span of the phonon branches from $\mathrm{Mg_3AsBr_3}$ to $\mathrm{Mg_3SbBr_3}$ and $\mathrm{Mg_3BiBr_3}$. Specifically, $\mathrm{Mg_3AsBr_3}$ exhibits the largest bandwidths (up to approximately 3.0~THz), while $\mathrm{Mg_3SbBr_3}$ and $\mathrm{Mg_3BiBr_3}$ show progressively narrower distributions, with the latter exhibiting the smallest overall dispersion. This trend reflects the increasing atomic mass of the pnictogen atom (As~$\rightarrow$~Sb~$\rightarrow$~Bi), which leads to a softening of the lattice vibrations and a corresponding reduction in phonon frequencies.

\begin{figure*}[tbhp]
\centering
\includegraphics[width=\textwidth]{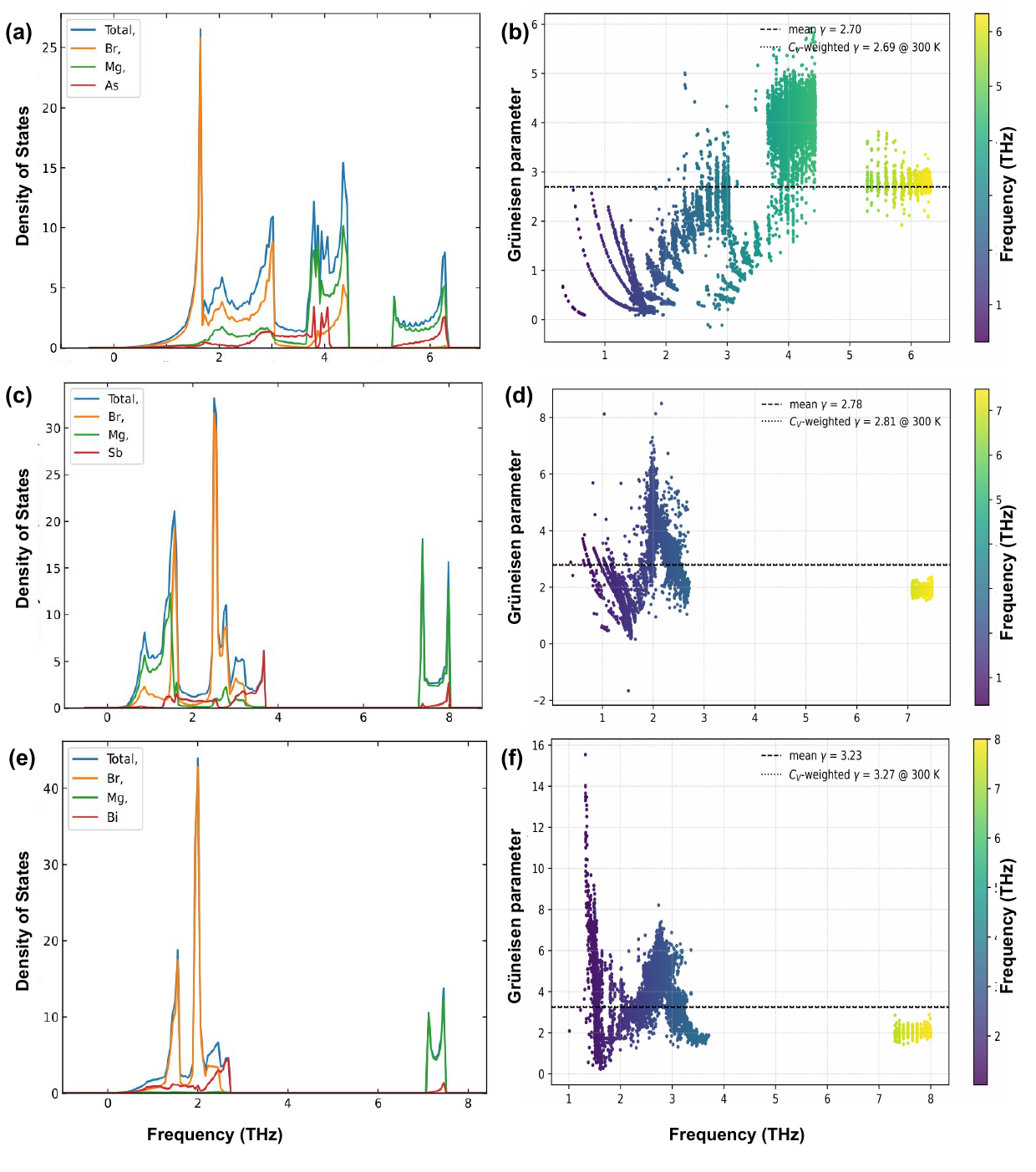}
\caption{Phonon DOS and PDOS of Mg$_3Z$Br$3$ (a) Z = As, (c) Z = Sb, (e) Z = Bi. And mode Gr\"{u}neisen  parameter $\gamma{\mathbf{q}\nu}$ vs.mode frequency for (b) Z = As, (d) X = Sb, (f) Z = Bi. In the Gr\"{u}neisen  plots, each point corresponds to a phonon mode (color-coded by frequency for clarity). The average Gr\"{u}neisen  parameter (dashed line) and the heat-capacity-weighted average at 300K (dotted line) are indicated.}
\label{fig:dos_gru}
\end{figure*}

Each phonon dispersion in Figure~\ref{fig:phonon_bands}(a,c and e) shows three acoustic modes (originating at $\Gamma$) and a manifold of optical modes at higher frequencies. There is a noticeable separation between the acoustic and optical phonon frequencies, especially pronounced for Mg$_3$SbBr$_3$ and Mg$_3$BiBr$_3$ compounds, where the highest acoustic mode ends below $\sim$2–4 THz and a gap appears before the optical modes around 3–4 THz. This partial phonon band gap arises because the heavy cation (Sb or Bi) and the Mg sublattice have much lower vibrational frequencies (acoustic and low optical modes), whereas the lighter atoms, especially Mg in all compounds, participate in the high-frequency optical modes as observed in Figure~\ref{fig:dos_gru}(a,c,e). The phonon density of states (DOS) in Figure~\ref{fig:dos_gru}(a,c,e) reflects this partitioning: the low-frequency region vibrations of the heavy $Z$ atoms is observed, while the high-frequency peaks correspond largely to motions of the lighter Mg and Br atoms\cite{Morelli2006}. For example, in Mg$_3$BiBr$_3$, the Bi atoms (the heaviest species) contribute primarily to the DOS below $\sim$3 THz (including the flat, low-energy optical modes and acoustic modes). In contrast, the sharp DOS features above 4 THz are mainly due to Mg–Br stretching motions with Bi remaining relatively immobile. A similar pattern is seen in Cs–Pb halide perovskites, where the heaviest cation (Cs) contributes to the lowest-frequency modes, and lighter anion vibrations dominate the high-frequency optical modes\cite{MorelliSlack2006-highk}. This distribution of phonon modes has implications for thermal properties: a large fraction of low-frequency modes (as in the Bi compound) means more modes are thermally occupied at a given temperature, boosting the heat capacity and entropy at lower $T$ compared to a system with fewer low-frequency modes.

Figure~\ref{fig:dos_gru} also presents the Gr\"{u}neisen  parameters $\gamma_{i}$ for each phonon mode (panels b, d, f), which measure the sensitivity of phonon frequency to volume changes. (By definition, $\gamma_i = -\frac{\partial \ln \omega_i}{\partial \ln V}$ for mode $i$, so modes that soften rapidly upon expansion have large positive $\gamma$.) We computed $\gamma_{i}$ for each mode via the quasi-harmonic approximation, using finite differences of phonon frequencies from volumes adjusted by $\pm1\%$, $\pm1.5\%$, and $\pm2\%$ around equilibrium. The mode Gr\"{u}neisen spectra (Figure~\ref{fig:dos_gru}b,d,f) reveal notably large Gr\"{u}neisen  values for the low-frequency branches, and the frequency spectrum average value $\bar{\gamma}$ is about 2.7 for Mg$_3$AsBr$_3$, 3.2 for Mg$_3$SbBr$_3$, and 2.8 for Mg$_3$BiBr$_3$ (dashed lines in Figure~\ref{fig:dos_gru}b,d,f). The heat-capacity-weighted averages are similarly high (e.g. $\gamma_{\mathit{Cv}} \approx 2.7$–3.3 at 300K, dotted lines in Figure~\ref{fig:dos_gru}b,d,f). These $\gamma$ values greatly exceed those of stiff, weakly anharmonic crystals (for instance, most rock-salt halides have $\gamma \approx 1.5$–1.9), indicating that Mg$_3X$Br$_3$ perovskites are strongly anharmonic. Physically, a large Gr\"{u}neisen  parameter signals that phonon frequencies are very sensitive to lattice expansion, which is a hallmark of soft bonding and pronounced anharmonicity (e.g., bonding involving stereochemically active lone pairs, as in Bi$^{3+}$, tends to produce giant anharmonic phonon softening\cite{MorelliSlack2006-highk}). The present compounds indeed have stereochemically active $ns^2$ cations (Sb$^{3-}$, Bi$^{3-}$ in the B-site position, formally analogous to Sb$^{3+}$/Bi$^{3+}$ with lone pairs) which likely contribute to the large $\gamma$ values via anharmonic rattling or off-centering tendencies.

Strong lattice anharmonicity and large Gr\"{u}neisen  parameters generally correlate with low lattice thermal conductivity, since phonon–phonon Umklapp scattering rates increase with $\gamma$ and phonon velocities decrease in soft lattices. According to Slack’s theory of lattice thermal conductivity, materials with high Debye temperature , low average $\gamma$, and simple crystal structures  will exhibit high thermal conductivity. The Mg$_3X$Br$_3$ perovskites violate these criteria – they have relatively low Debye temperatures (especially for Z = Bi), large Gr\"{u}neisen  constants $\gamma \sim 3$, and a complex 7-atom primitive cell ($n=7$) – and thus can be expected to possess quite low lattice thermal conductivities. Qualitatively, a high $\gamma$ reduces the phonon mean free path via strong anharmonic scattering, and the small sound velocities in these soft, heavy-mass lattices further limit heat transport\cite{Haeger2020-halide-thermal}. This is consistent with the behavior of other halide perovskites: for example, CsPbBr$_3$ single crystals have $\kappa_L \approx 0.4$–0.5~W,m$^{-1}$,K$^{-1}$ at room temperature, and Cs$_2$AgBiBr$_6$ was recently reported to have an ultralow $\kappa_L \sim 0.2$W,m$^{-1}$,K$^{-1}$\cite{Zheng2024-Cs2AgBiBr6}. We therefore expect that Mg$_3$SbBr$_3$ and Mg$_3$BiBr$_3$ will exhibit extremely low thermal conductivities, making them attractive for thermal barrier coatings or thermoelectric applications where heat flow suppression is desired \cite{Khandaker2025-Mg3AB3}.
 
\begin{figure}[tbhp]
\centering
\includegraphics[width=\textwidth]{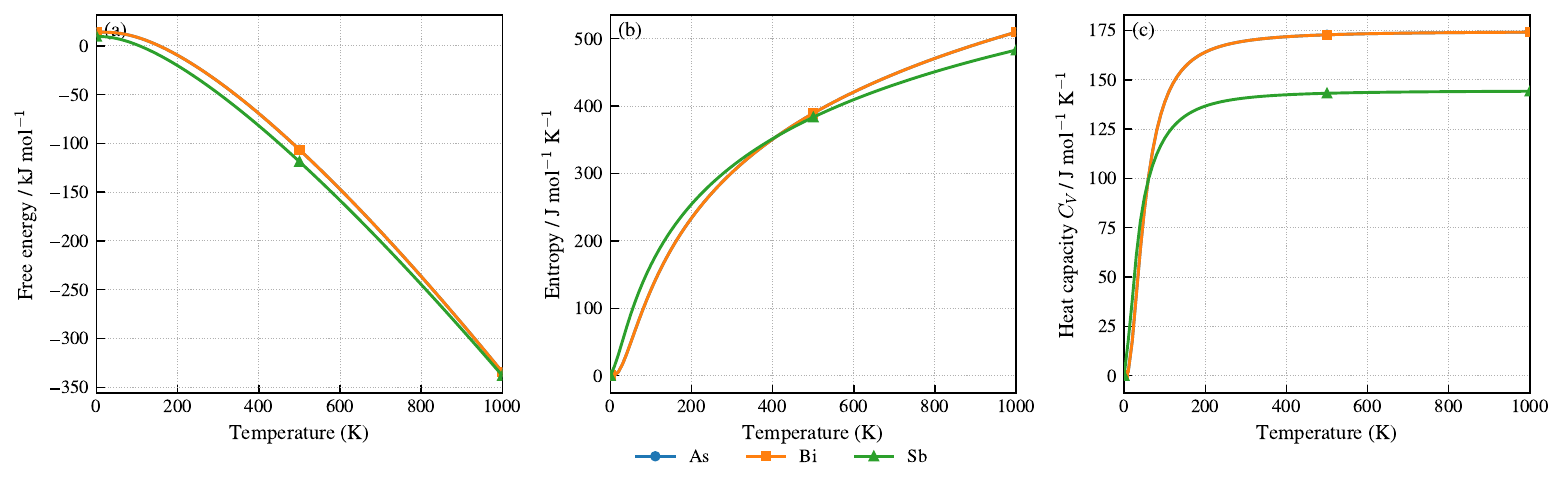}
\caption{Thermodynamic functions of Mg$_3Z$Br$_3$ (Z = As, Sb, Bi) obtained from phonon calculations; (a) molar Helmholtz free energy $F$, (b) entropy $S$, and (c) constant-volume heat capacity $C_V$, as a function of temperature.}
\label{fig:thermal}
\end{figure}

Finally, the thermodynamic properties obtained from the phonon calculations are summarized in Figure~\ref{fig:thermal}. The temperature dependence of the Helmholtz free energy $F(T)$, entropy $S(T)$ and constant-volume molar heat capacity $C_V(T)$ is shown for Z = As, Sb, and Bi in Figure~\ref{fig:thermal} (a) to (c), respectively. All three compounds obey the expected Debye–Einstein trends. At low temperatures ($T \ll \theta_D$), $C_V$ rises as $\sim T^3$ and $S \sim T^3$ (as indicated by the initial slopes in Figure~\ref{fig:thermal}). Free energy curves $F(T)$ in Figure~\ref{fig:thermal} (a) show Mg$_3$SbBr$_3$ having the lowest vibrational free energy (most negative) at low $T$, though $F(T)$ converges among the compounds at higher temperature as $-TS$ dominates. By 1000K, the entropy of Mg$_3$BiBr$_3$ reaches on the order of $500\text{J mol}^{-1}\text{K}^{-1}$, while Mg$_3$SbBr$_3$ is around $480~\text{J mol}^{-1}\text{K}^{-1}$ (Figure~\ref{fig:thermal} - b). Around room temperature, $C_V$ approaches a substantial fraction of the Dulong–Petit limit. Notably, Mg$_3$AsBr$3$ and Mg$_3$BiBr$3$ has a larger heat capacity at 300K than Mg$_3$SbBr$_3$ as evident from Figure~\ref{fig:thermal} (c). At $T \gtrsim 300$K, the heat capacity $C_V(T)$ of each compound begins to saturate near the Dulong–Petit value, and correspondingly $S(T)$ grows more slowly.
In summary, the lattice dynamics of Mg$_3Z$Br$_3$ are characterized by: (i) dynamic stability of the cubic perovskite structure (no soft-mode instabilities at 0~K for Z = As, Sb; while instability for Z = Bi may exist, correlating with Bi’s low Goldschmidt tolerance factor), (ii) a systematic softening of the phonon spectrum from As to Bi due to increasing atomic mass and lattice size, (iii) phonon modes divided into low-frequency (acoustic and low optical) modes mainly involving heavy cation motion and high-frequency modes involving lighter atoms, (iv) unusually large Grüneisen parameters indicating strong anharmonicity. These findings highlight the important role of the Goldschmidt tolerance factor and atomic masses in governing the vibrational stability of halide perovskites, as well as the interplay between phonon dispersion and anharmonicity in determining thermal transport. Future work should explore finite-temperature lattice dynamics (e.g., quartic anharmonic stabilization of any soft modes and high-temperature phase transitions) and possibly measure the thermal conductivity of these materials to validate the ultra-low values predicted (on par with the most thermally resistive perovskites like Cs$_2$AgBiBr$_6$ \cite{Zheng2024-Cs2AgBiBr6}).

\section{Electronic Properties and Band Structure Analysis}

DFT calculations yield indirect gaps for all three compounds with a monotonic decrease across the pnictogen series (Table~\ref{tab:electronic_effmass_merged}): $E_\mathrm{g}^{\mathrm{HSE}}$ falls from 2.0645~eV (Mg$_3$AsBr$_3$) to 1.6533~eV (Mg$_3$SbBr$_3$) to 1.5226~eV (Mg$_3$BiBr$_3$). This decrease of band gap aligns with the argument of Kim and co-authors  for InP/InAs/InSb using hybrid functionals~\cite{kim2009}. Modulation of lone-pair stereochemical activity in Bi compounds provides further support for the decrease of band gap~\cite{ogawa2024}. A similar down-Group~15 sequence is reported for NaZnX (X = P, As, Sb, Bi) associated with nontrivial band topology~\cite{Lee2022}. The band-gap range (1.5–2.1~eV) suits single-junction PV; Mg$_3$BiBr$_3$ (1.52~eV) and Mg$_3$SbBr$_3$ (1.65~eV) fall within the Shockley–Queisser limit~\cite{Shockley1961}, although the indirect gap favors thicker absorbers. 

\begin{table}[htbp]
\centering
\caption{Electronic band gap, carrier mobilities, and direction-resolved effective masses of Mg$_3$ZBr$_3$ (Z = As, Sb, Bi).}
\label{tab:electronic_effmass_merged}
\setlength{\tabcolsep}{6pt}
\begin{tabular}{@{} l
                  r                    
                  r                    
                  r                    
                  r                    
                  l                    
                  S[table-format=1.3]  
                  S[table-format=-1.3] 
                  @{}}
\toprule
\multicolumn{1}{c}{Material} &
\multicolumn{1}{c}{$E_\mathrm{g}^{\mathrm{HSE}}$} &
\multicolumn{1}{c}{$E_\mathrm{g}^{\mathrm{PBE}}$} &
\multicolumn{1}{c}{$\mu_n$} &
\multicolumn{1}{c}{$\mu_p$} &
\multicolumn{1}{c}{$k$-path} &
\multicolumn{1}{c}{$m_e^\ast/m_0$} &
\multicolumn{1}{c}{$m_h^\ast/m_0$} \\
\multicolumn{1}{c}{} &
\multicolumn{1}{c}{(eV)} &
\multicolumn{1}{c}{(eV)} &
\multicolumn{1}{c}{(cm$^2$V$^{-1}$s$^{-1}$)} &
\multicolumn{1}{c}{(cm$^2$V$^{-1}$s$^{-1}$)} &
\multicolumn{1}{c}{} &
\multicolumn{1}{c}{} &
\multicolumn{1}{c}{} \\
\midrule
\multirow{2}{*}{Mg$_3$AsBr$_3$}
  & \multirow{2}{*}{2.0645} & \multirow{2}{*}{1.2907} & \multirow{2}{*}{57.86} & \multirow{2}{*}{57.86} & $R \rightarrow \Gamma$ & 0.336 & -1.096 \\
  &                          &                        &                        &                        & $\Gamma \rightarrow R$ & 0.291 & -0.959 \\
\addlinespace[2pt]
\multirow{2}{*}{Mg$_3$SbBr$_3$}
  & \multirow{2}{*}{1.6533} & \multirow{2}{*}{0.9259}  & \multirow{2}{*}{57.86} & \multirow{2}{*}{30.22} & $R \rightarrow \Gamma$ & 0.335 & -0.989 \\
  &                          &                        &                        &                        & $\Gamma \rightarrow R$ & 0.253 & -0.648 \\
\addlinespace[2pt]
\multirow{2}{*}{Mg$_3$BiBr$_3$}
  & \multirow{2}{*}{1.5226} & \multirow{2}{*}{0.8587}  & \multirow{2}{*}{57.86} & \multirow{2}{*}{59.20} & $R \rightarrow \Gamma$ & 0.335 & -0.950 \\
  &                          &                        &                        &                        & $\Gamma \rightarrow R$ & 0.238 & -0.565 \\
\bottomrule
\end{tabular}
\end{table}

\begin{figure}
    \centering
    \includegraphics[width=1.0\linewidth]{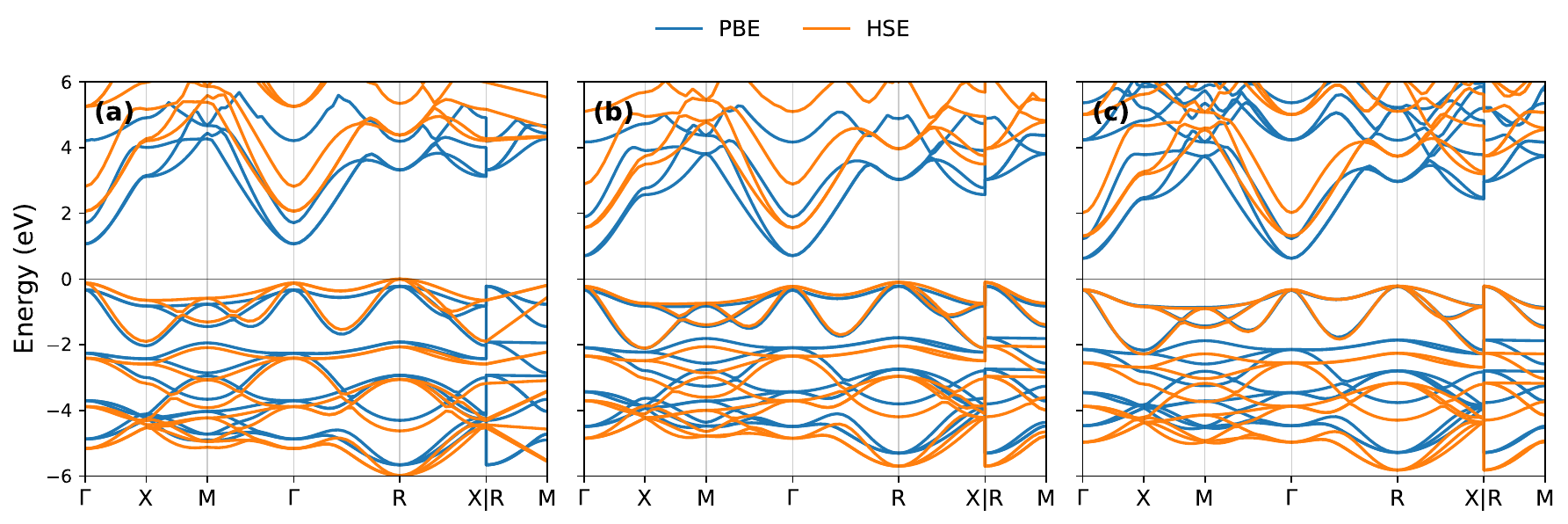}
    \caption{
    Electronic band structures of (a)~Mg\textsubscript{3}AsBr\textsubscript{3}, 
    (b)~Mg\textsubscript{3}SbBr\textsubscript{3}, and (c)~Mg\textsubscript{3}BiBr\textsubscript{3} computed with PBE (blue dashed) and HSE06 (red solid); the horizontal dotted line marks the Fermi level at $E=0$~eV.
    The high‐symmetry path is $\Gamma$–X–M–$\Gamma$–R–X$\mid$R–M.
    The fundamental gaps are
    \(E_\mathrm{g}^{\mathrm{HSE}} = 2.0645\,\mathrm{eV}
      \bigl(E_\mathrm{g}^{\mathrm{PBE}} = 1.2907\,\mathrm{eV}\bigr)\)
    for (a),
    \(E_\mathrm{g}^{\mathrm{HSE}} = 1.6533\,\mathrm{eV}
      \bigl(E_\mathrm{g}^{\mathrm{PBE}} = 0.9259\,\mathrm{eV}\bigr)\)
    for (b), and
    \(E_\mathrm{g}^{\mathrm{HSE}} = 1.5226\,\mathrm{eV}
      \bigl(E_\mathrm{g}^{\mathrm{PBE}} = 0.8587\,\mathrm{eV}\bigr)\)
    for (c).
    }
    \label{fig:band}
\end{figure}

Effective masses were obtained from the local band curvature,
\begin{equation}
m^* = \hbar^2 \left( \frac{d^2E}{dk^2} \right)^{-1},
\end{equation}
using fits near the band edges. The values appear in Table~\ref{tab:electronic_effmass_merged}. All compounds have shown flat VBM dispersion (large $|m_h^\ast|$) and a more dispersive CBM at $\Gamma$ (smaller $m_e^\ast$) (Figure~\ref{fig:band})~\cite{williamson2017}. This asymmetry points to higher electron than hole mobility, consistent with relations between dispersion and transport~\cite{hautier}.The mobility was calculated using the following equation-2 where, $\mu$ denotes the carrier mobility (in cm$^{2}$\,V$^{-1}$\,s$^{-1}$), $C$ represents the longitudinal elastic modulus (such as $C_{11}$). The symbol $m^{\ast}$ denotes the carrier effective mass. Finally, $E_{1}$ corresponds to the longitudinal deformation potential, and the calculated values are listed in Table-\ref{tab:params}.

\begin{equation}
\mu = \frac{\sqrt{8\pi}\, e\, \hbar^{4}\, C}{3\, (m^{\ast})^{5/2}\, (k_{\mathrm B} T)^{3/2}\, E_{1}^{2}}
\end{equation}

The projected DOS identifies the orbital contributions in the electronic DOS. In all three compounds, the valence band is dominated by Br~4$p$ states with increasing pnictogen $p$ character near the VBM; Mg's contribution near the edges is small. The fundamental gap is a transition from Br~4$p$–dominated valence states to conduction states with pnictogen $s/p$ and Mg~$s$ orbital, consistent with prior perovskite studies~\cite{Yin2014, Filip2014}. The indirect gaps affect optical and electronic response: indirect absorption is weaker (phonon assistance required), while nonradiative recombination can be reduced, extending lifetimes.

\begin{figure}
    \centering
    \includegraphics[width=1.0\linewidth]{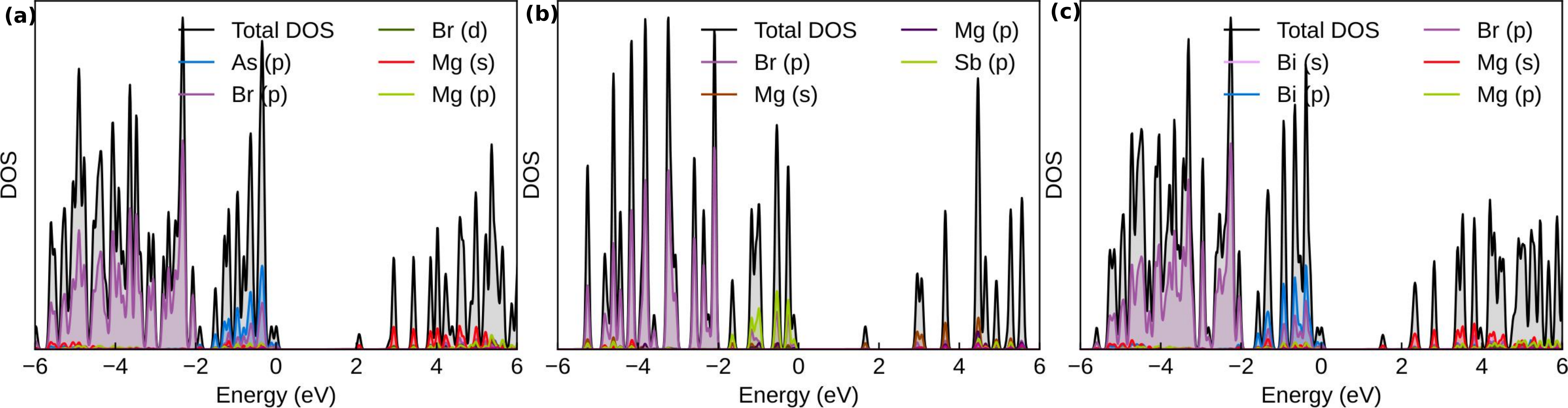}
    \caption{Electronic DOS of  (a) Mg$_3$AsBr$_3$, (b) Mg$_3$BiBr$_3$, (c) Mg$_3$SbBr$_3$ with HSE06 functional. Element-projected contributions follow the legend; total DOS in black.}
    \label{fig:DOS}
\end{figure}

\begin{figure}
    \centering
    \includegraphics[width=1\linewidth]{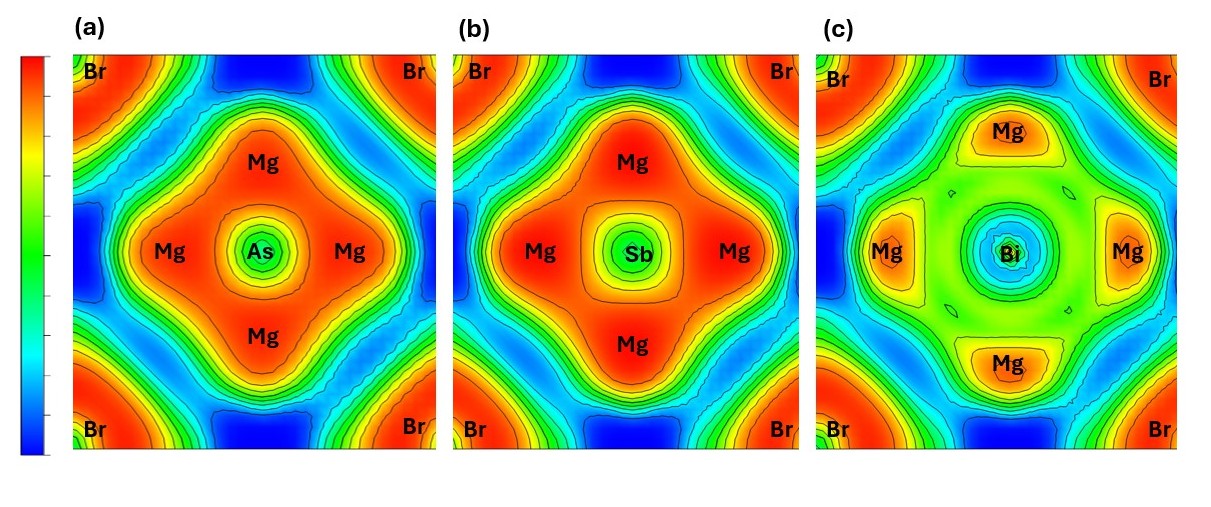}
    \caption{Two-dimensional ELF maps in the (100) plane of
    (a)~\ce{Mg3AsBr3}, (b)~\ce{Mg3SbBr3}, and (c)~\ce{Mg3BiBr3}.
    Colour scale: ELF = 0 (blue) to 1 (red). }
    \label{fig:elf}
\end{figure}

Figure~\ref{fig:elf} shows the (100)-plane electron localization function (ELF) of \ce{Mg3AsBr3}, \ce{Mg3SbBr3}, and \ce{Mg3BiBr3}. All three exhibit a pronounced $ns^2$ lone-pair basin at the pnictogen. The maximum ELF value decreases from As  to Sb  to Bi as observed from Figure~\ref{fig:elf} (a) - (c) . Here outside the lone-pair basin, Mg--Br linkages have ELF$<0.2$ (ionic), whereas Pn--Br bond lines retain intermediate ELF ($\sim$0.45–0.60), indicating mixed ionic–covalent character that strengthens from As to Bi due to increased Pn~$p$/$s$--Br~$p$ overlap\cite{Varadwaj2022}. Bonding explains the band gap trends as well. The Mg--Br network is largely ionic, while Pn--Br bonds show mixed character. From As to Bi, stronger Pn~$p$/$s$--Br~$p$ hybridization increases antibonding weight at the CBM and narrows $E_g$. The stereochemically active $ns^2$ lone pairs on Sb and Bi add states near the band edges, a behavior well documented in related Bi compounds~\cite{Walsh2011}.

\section{Optical Properties and Spectroscopic Analysis}

 The computed optical parameters include the complex dielectric function $\varepsilon(\omega) = \varepsilon_1(\omega) + i\varepsilon_2(\omega)$, absorption coefficient $\alpha(\omega)$, reflectivity $R(\omega)$, refractive index $n(\omega)$, and extinction coefficient $\kappa(\omega)$. These properties are fundamentally connected to the electronic structure and provide critical insights into the light-matter interactions essential for applications in photovoltaics, photodetectors, and other optoelectronic devices.

\begin{figure}
    \centering
    \includegraphics[width=1\linewidth]{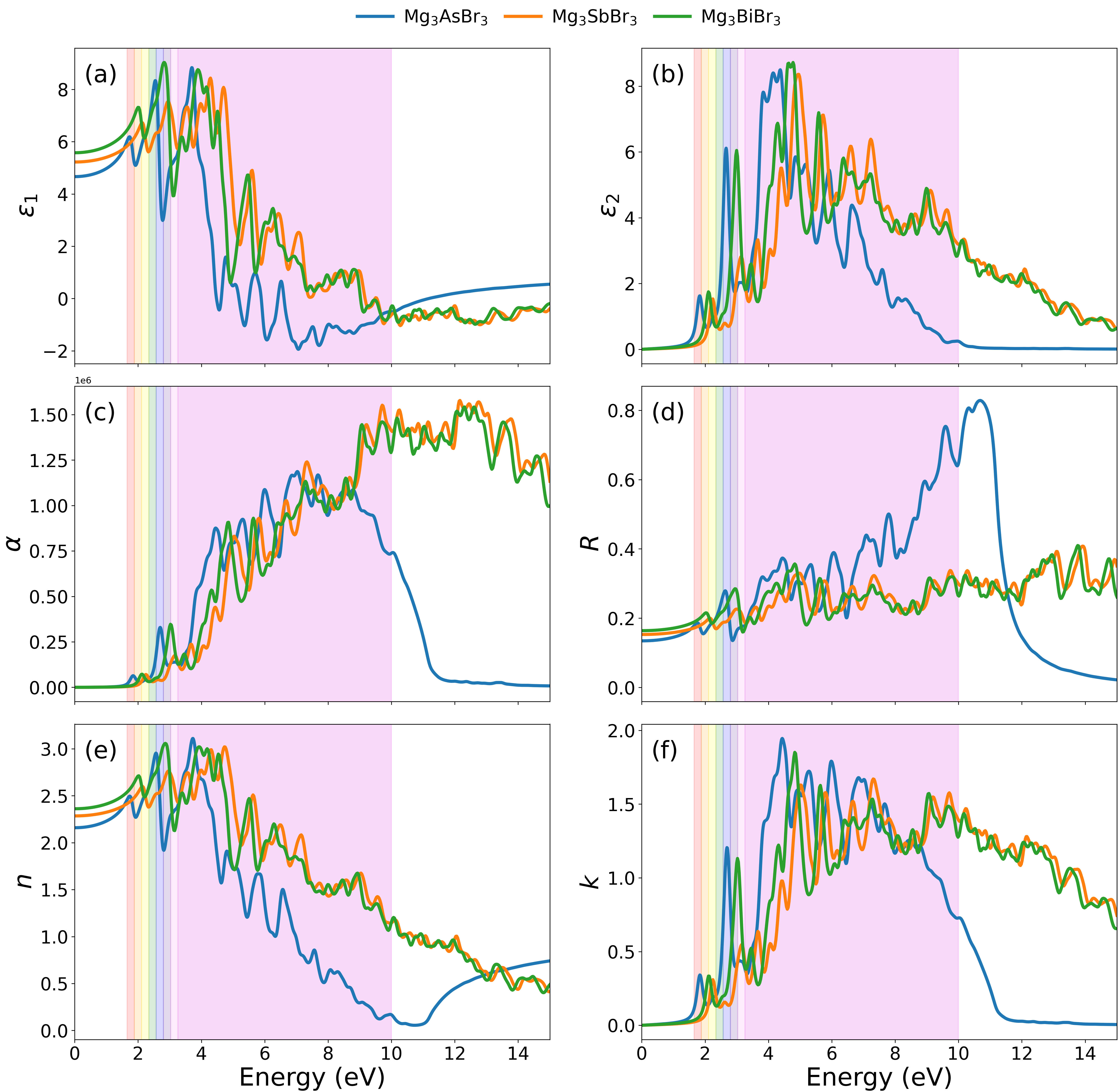}
\caption{Calculated optical properties of Mg$_3$ZBr$_3$ (Z = As, Sb, Bi) perovskites as a function of photon energy: (a) real part of the dielectric function, (b) imaginary part of the dielectric function, (c) optical absorption coefficient, (d) optical reflectivity, (e) refractive index, and (f) extinction coefficient. The colored vertical bands in the low-energy region  represent the visible region, while the violet-shaded regions highlight the UV spectrum. } 
\label{fig:optical_property}
\end{figure}

Figure\ref{fig:optical_property} (a) and (b ) display both the real ($\varepsilon_1$) and imaginary ($\varepsilon_2$) components of the dielectric function for all three compounds across the 0-15 eV energy range. The static dielectric constants ($\varepsilon_1(0)$) for the three materials were determined to be approximately 4.8, 5.7, and 5.9 for Mg$_3$AsBr$_3$, Mg$_3$SbBr$_3$, and Mg$_3$BiBr$_3$, respectively. This systematic increase with heavier pnictogen elements correlates with the decreasing band gap is consistent with the Penn model \cite{RAVINDRA200721} \cite{Penn1962}.

 Notably, in the real part as in Figure 9(a), all three materials show pronounced peaks in the 2-7 eV region, corresponding to critical points in their joint density of states where significant interband transitions occur. Mg$_3$AsBr$_3$ displays sharper spectral features with more distinct oscillations compared to the other two compounds, which exhibit somewhat broadened spectral features. The dielectric function approaches asymptotic values at higher energies ($>$12 eV), indicating transitions to the free-electron-like regime. The imaginary part $\varepsilon_2(\omega)$ as in Figure 9(b), which directly relates to the absorption of photons, exhibits significant absorption peaks after 2 eV range. The onset of absorption is  consistent with the calculated band gaps: $\sim$2.06 eV for Mg$_3$AsBr$_3$, $\sim$1.65 eV for Mg$_3$SbBr$_3$, and $\sim$1.52 eV for Mg$_3$BiBr$_3$. Both Mg$_3$SbBr$_3$ and Mg$_3$BiBr$_3$ display somewhat similar spectral profiles. Based on the projected density of states analysis, the primary transitions can be attributed to Br 4p $\rightarrow$ pnictogen p/s transitions, with the exact energy dependent on the specific pnictogen element \cite{Yu2010}.

The absorption coefficient $\alpha(\omega)$, calculated from the complex dielectric function using:

\begin{equation}
\alpha(\omega) = \frac{2\omega}{c} \sqrt{\frac{\sqrt{\varepsilon_1^2(\omega) + \varepsilon_2^2(\omega)} - \varepsilon_1(\omega)}{2}}
\end{equation}

This provides a quantitative measure of the light absorption capability of the materials. As shown in Figure-\ref{fig:optical_property} (c), all three compounds exhibit strong absorption in the visible and ultraviolet regions, with absorption coefficients reaching approximately 0.8$\times$10$^6$-1.5$\times$10$^6$ cm$^{-1}$ at higher photon energies. These characteristics align with the indirect band gap nature of these materials, where phonon-assisted transitions are necessary for photon absorption near the band edge, resulting in relatively weak absorption near the onset. However, the absorption coefficients increase rapidly at energies well above the band gap, where direct transitions become dominant \cite{Yu2010}. The reflectivity $R(\omega)$, calculated from the complex refractive index using:

\begin{equation}
R(\omega) = \left|\frac{\sqrt{\varepsilon(\omega)} - 1}{\sqrt{\varepsilon(\omega)} + 1}\right|^2
\end{equation}

This represents the fraction of incident light reflected at the material surface. The reflectivity spectra in Figure-\ref{fig:optical_property} (d) show distinctive features that correlate with the dielectric function. In the low-energy region (near the band gap), the reflectivity values are approximately 0.17, 0.185, and 0.19 for Mg$_3$AsBr$_3$, Mg$_3$SbBr$_3$, and Mg$_3$BiBr$_3$, respectively, indicating that 17-19\% of incoming light is reflected from the surface. This relatively low reflectivity in the visible range  is advantageous for maximizing light coupling into the absorber layer \cite{Green2014}. The reflectivity increases significantly in regions of strong absorption, reaching maxima of approximately 0.82 for Mg$_3$AsBr$_3$ around 11 eV, while Mg$_3$SbBr$_3$ and Mg$_3$BiBr$_3$ show lower maximum reflectivity values of approximately 0.35-0.40 in the high-energy region. A particularly interesting feature is the significantly higher reflectivity of Mg$_3$AsBr$_3$ compared to the other compounds in the 9-12 eV range, which suggests fundamentally different high-energy electronic transitions. This difference may be attributed to the distinct orbital character and energy positioning of the As 4p states compared to the heavier pnictogens' p states \cite{Pauling1960}. The refractive index $n(\omega)$, derived from the real part of the dielectric function as:

\begin{equation}
n(\omega) = \sqrt{\frac{\sqrt{\varepsilon_1^2(\omega) + \varepsilon_2^2(\omega)} + \varepsilon_1(\omega)}{2}}
\end{equation}

This provides information about light propagation through the material, which is represented in Figure~\ref{fig:optical_property} (e). The calculated static refractive indices (at zero frequency) are approximately 2.2, 2.4, and 2.45 for Mg$_3$AsBr$_3$, Mg$_3$SbBr$_3$, and Mg$_3$BiBr$_3$, respectively. This trend follows the Moss relation that $n^4 \cdot E_g$ is approximately constant across similar material families \cite{Moss1985}. The moderate refractive indices (2.2-2.5) of the studied materials in the near-infrared region could be exploited for optical coating applications. The refractive index spectra show significant dispersion in the visible and ultraviolet regions, with maximum values of approximately 3.1, 3.0, and 3.0 for the As, Sb, and Bi compounds, respectively, occurring in the 3-5 eV range. At energies above 8 eV, the refractive index decreases substantially for all three materials, approaching values below 1 in regions of strong absorption, a phenomenon associated with anomalous dispersion near resonant absorption frequencies \cite{Fox2010}. The extinction coefficient $\kappa(\omega)$, which represents the imaginary part of the complex refractive index and is directly related to the absorption coefficient through:

\begin{equation}
\kappa(\omega) = \frac{\alpha(\omega)\lambda}{4\pi}
\end{equation}

 This provides a measure of the attenuation of light as it passes through the material. The extinction coefficient spectra closely mirror the trends observed in the imaginary part of the dielectric function. All three compounds in Figure-\ref{fig:optical_property} (f) show maximum extinction coefficients in the 3-7 eV range, with peak values of approximately 1.95, 1.65, and 1.60 for Mg$_3$AsBr$_3$, Mg$_3$SbBr$_3$, and Mg$_3$BiBr$_3$, respectively. The extinction coefficient of Mg$_3$AsBr$_3$ drops to nearly zero above 12 eV, while the heavier pnictogen compounds maintain moderate extinction coefficients even at higher energies, suggesting different high-energy absorption mechanisms possibly related to deeper core-level transitions \cite{Yu2010}.

\section{Elastic Properties and Mechanical Stability}

The elastic properties of materials are fundamental characteristics that describe their mechanical response to applied forces within the reversible deformation regime. These properties are essential for understanding structural stability, mechanical behavior, and potential applications in various technological domains. The investigated materials crystallize in the cubic perovskite structure with space group Pm-3m (\#221), exhibiting high symmetry that reduces the independent elastic constants to only three: $C_{11}$, $C_{12}$, and $C_{44}$. Where $C_{11}$ represents longitudinal deformation resistance, $C_{12}$ describes the cross-coupling between perpendicular deformations, and $C_{44}$ represents the resistance to shear deformation. Table \ref{tab:merged_mechanical_swapped} presents the calculated elastic constants for all three studied compounds. The elastic constants display a clear trend, with all values decreasing from Mg$_3$AsBr$_3$ to Mg$_3$BiBr$_3$, which can be attributed to the increasing atomic radius of the pnictogen element and the consequent expansion of the lattice parameters. This observation aligns with the well-established principles of bond length-strength correlation in crystalline solids, where longer bonds typically exhibit weaker bond strength, leading to reduced elastic constants. A similar trend has been observed in various perovskite systems, as reported by Fadda et al. \cite{Ekuma} and Savinov et al. \cite{Savinov2008}.

The mechanical stability of a crystalline material is governed by the Born stability criteria, which for cubic crystals require that: $C_{11} - C_{12} > 0$ (resistance to tetragonal shear deformation). $C_{11} + 2C_{12} > 0$ (resistance to uniform volume compression) and $C_{44} > 0$ (resistance to trigonal shear deformation).
As evident from Table \ref{tab:merged_mechanical_swapped}, Mg$_3$AsBr$_3$, Mg$_3$SbBr$_3$, and Mg$_3$BiBr$_3$ satisfy all these criteria, confirming their mechanical stability. The positive eigenvalues of the stiffness matrix further corroborate this conclusion. The value of $C_{11} - C_{12}$, is 61.  038 GPa for Mg$_3$AsBr$_3$ , 54.108 GPa for Mg$_3$SbBr$_3$, and 50.105 GPa for Mg$_3$BiBr$_3$ indicate substantial resistance to tetragonal deformations, which is important for maintaining structural integrity under anisotropic stress conditions. The ratio $C_{12}/C_{11}$, which represents the Cauchy pressure normalized by $C_{11}$, provides insights into the nature of interatomic bonding. The calculated values of the Cauchy pressure are 0.277 for Mg$_3$AsBr$_3$, 0.276 for Mg$_3$SbBr$_3$, and 0.272 for Mg$_3$BiBr$_3$ are relatively low, suggesting a significant degree of directional covalent character in their bonding. This observation is consistent with the negative Cauchy pressure ($C_{12} - C_{44}$) values of -4.6 GPa, -1.9 GPa and -1.7 GPa for Mg$_3$AsBr$_3$, Mg$_3$SbBr$_3$ and Mg$_3$BiBr$_3$ respectively as mentioned in Table \ref{tab:merged_mechanical_swapped}, which further indicating predominant covalent bonding characteristics with some ionic contribution, as mentioned by Pettifor \cite{Pettifor1992} and others in their seminal works on the relationship between Cauchy pressure and bonding nature. This is consistent with the results found in ELF calculations of Figure-\ref{fig:elf}. From the elastic constants, various important mechanical parameters can be derived, including bulk modulus ($B$), Young's modulus ($E$), shear modulus ($G$), and Poisson's ratio ($\nu$). For polycrystalline materials, these properties are typically estimated using the Voigt, Reuss, and Hill approximations, which provide upper-bound, lower-bound, and average estimates, respectively. Table \ref{tab:merged_mechanical_swapped} summarizes the calculated Hill average values of these properties for the Mg$_3$ZBr$_3$ compounds.

\begin{table}[htbp]
\centering
\scriptsize
\setlength{\tabcolsep}{3pt}
\caption{Hill-averaged polycrystalline mechanical parameters and principal
         single–crystal elastic constants of
         Mg\textsubscript{3}ZBr\textsubscript{3} (Z=As, Sb, Bi).}
\resizebox{\textwidth}{!}{%
\begin{tabular}{lccccccccccccc}
\toprule
 & \multicolumn{5}{c}{\textbf{Elastic constants (GPa)}} &
   \multicolumn{8}{c}{\textbf{Polycrystalline moduli and derived parameters}}\\
\cmidrule(r){2-6}\cmidrule(l){7-14}
Compound &
$C_{11}$ & $C_{12}$ & $C_{44}$ & $C_{11}\!-\!C_{12}$ & $C_{11}\!+\!2C_{12}$ &
$B$ & $E$ & $G$ & $H_\mathrm{V}$ & $\nu$ & $B/G$ & $\zeta$ & $\Theta_{\text D}$\,(K) \\
\midrule
Mg$_3$AsBr$_3$ & 84.485 & 23.447 & 28.033 & 61.038 & 131.379 & 43.793 & 71.273 & 29.002 & 3.323 & 0.229 & 1.510 & 0.919 & 314.8 \\
Mg$_3$SbBr$_3$ & 74.694 & 20.586 & 22.460 & 54.108 & 115.866 & 38.622 & 60.050 & 24.197 & 2.814 & 0.241 & 1.596 & 0.830 & 276.6 \\
Mg$_3$BiBr$_3$ & 68.824 & 18.717 & 20.448 & 50.105 & 106.258 & 35.419 & 55.275 & 22.181 & 2.598 & 0.240 & 1.597 & 0.816 & 242.9 \\
\bottomrule
\end{tabular}}
\label{tab:merged_mechanical_swapped}
\end{table}

The bulk modulus ($B$) represents the material's resistance to hydrostatic compression and is inversely proportional to the compressibility. The calculated value of bulk moduli of 43.793 GPa for Mg$_3$AsBr$_3$, 38.622 GPa for Mg$_3$SbBr$_3$ and 35.419 GPa for Mg$_3$BiBr$_3$ are relatively moderate compared to typical oxide perovskites like BaTiO$_3$ ($\sim$162 GPa) or SrTiO$_3$ ($\sim$179 GPa), but higher than many halide perovskites such as CsPbI$_3$ ($\sim$19.8 GPa) as reported by Rakita et al. \cite{Rakita2017}. The observed decrease in bulk modulus from Mg$_3$AsBr$_3$ to Mg$_3$BiBr$_3$ can be attributed to the increased lattice parameter and subsequently weaker average bond strength. The bulk modulus can be related to the interatomic bonding strength through empirical relationships such as the one proposed by Cohen \cite{PhysRevB_cohen}:

\begin{equation}
B = k \cdot d^{-3.5}
\end{equation}

where $d$ is the average bond length, and $k$ is a proportionality constant. The decreasing bulk modulus with increasing pnictogen size follows this relationship, indicating the dominant role of bond length in determining compressibility in these materials.

The Young's modulus ($E$) represents the material's stiffness under uniaxial deformation, while the shear modulus ($G$) describes resistance to shape change under shear stress. Both moduli show a decreasing trend from Mg$_3$AsBr$_3$ to Mg$_3$BiBr$_3$. This reduction is more pronounced than that observed for the bulk modulus, suggesting that the materials become particularly less resistant to directional (shear and tensile) deformations as the pnictogen size increases, a behavior that can be attributed to the greater polarizability of larger atoms leading to more easily distorted electron clouds under directional stress \cite{Burdett1982}. The Poisson's ratio ($\nu$), which represents the negative ratio of transverse to axial strain under uniaxial stress, provides valuable insights into the nature of interatomic forces. As seen from Table- \ref{tab:merged_mechanical_swapped} calculated values are within the typical range for ionic and covalent solids (0.2-0.3). According to the classification scheme proposed by Greaves et al. \cite{Greaves2011}, materials with Poisson's ratios below 0.25 tend to exhibit more open structures with predominantly directional bonding, aligning with the covalent character suggested by the negative Cauchy pressure. The brittleness or ductility of materials can be assessed using Pugh's ratio ($B/G$), with a critical value of 1.75 typically distinguishing between brittle ($<$1.75) and ductile ($>$1.75) behavior. The calculated Pugh's ratios of 1.510 for Mg$_3$AsBr$_3$, 1.596 for Mg$_3$SbBr$_3$, and 1.597 for Mg$_3$BiBr$_3$ suggest that all materials exhibit brittle characteristics, although the higher values indicate a slight trend toward increased ductility with increasing pnictogen size. This trend can be rationalized by considering that the larger, more polarizable Sb and Bi atoms allow for greater structural flexibility and plastic deformation before bond breaking \cite{Pugh1954}. The brittle nature indicated by Pugh's ratios below 1.75 does necessitate careful handling during device fabrication to prevent fracture \cite{Green2014}. The Cauchy pressure ($C_{12} - C_{44}$) provides an additional perspective on ductility, with negative values typically associated with directional bonding and brittle behavior, while positive values indicate more delocalized metallic bonding and ductility. The negative Cauchy pressures of these materials corroborate the brittle nature of these materials while suggesting a slight trend toward increased metallicity \cite{Pettifor1992}. The Debye temperature ($\Theta_D$), calculated from the average sound velocity using the quasi-harmonic Debye model, provides insights into lattice thermal properties and phonon behavior. The estimated values that are mentioned in Table- \ref{tab:merged_mechanical_swapped} are relatively low compared to typical oxide perovskites, indicating softer lattice phonons and potentially higher phonon scattering rates. This characteristic could be beneficial for thermoelectric applications where low thermal conductivity is desirable. The decrease in Debye temperature from Mg$_3$AsBr$_3$ to Mg$_3$BiBr$_3$ is consistent with the trend in elastic moduli and can be attributed to the lower frequency vibrations associated with the heavier pnictogen atoms and weaker interatomic forces \cite{Anderson1963}. The trend of decreasing elastic moduli and Debye temperature from Mg$_3$AsBr$_3$ to Mg$_3$BiBr$_3$ suggests that the Mg$_3$BiBr$_3$ compound might exhibit even lower thermal conductivity, potentially beneficial for thermoelectric performance \cite{Snyder2008}.

\begin{figure}
    \centering
    \includegraphics[width=1.0\linewidth]{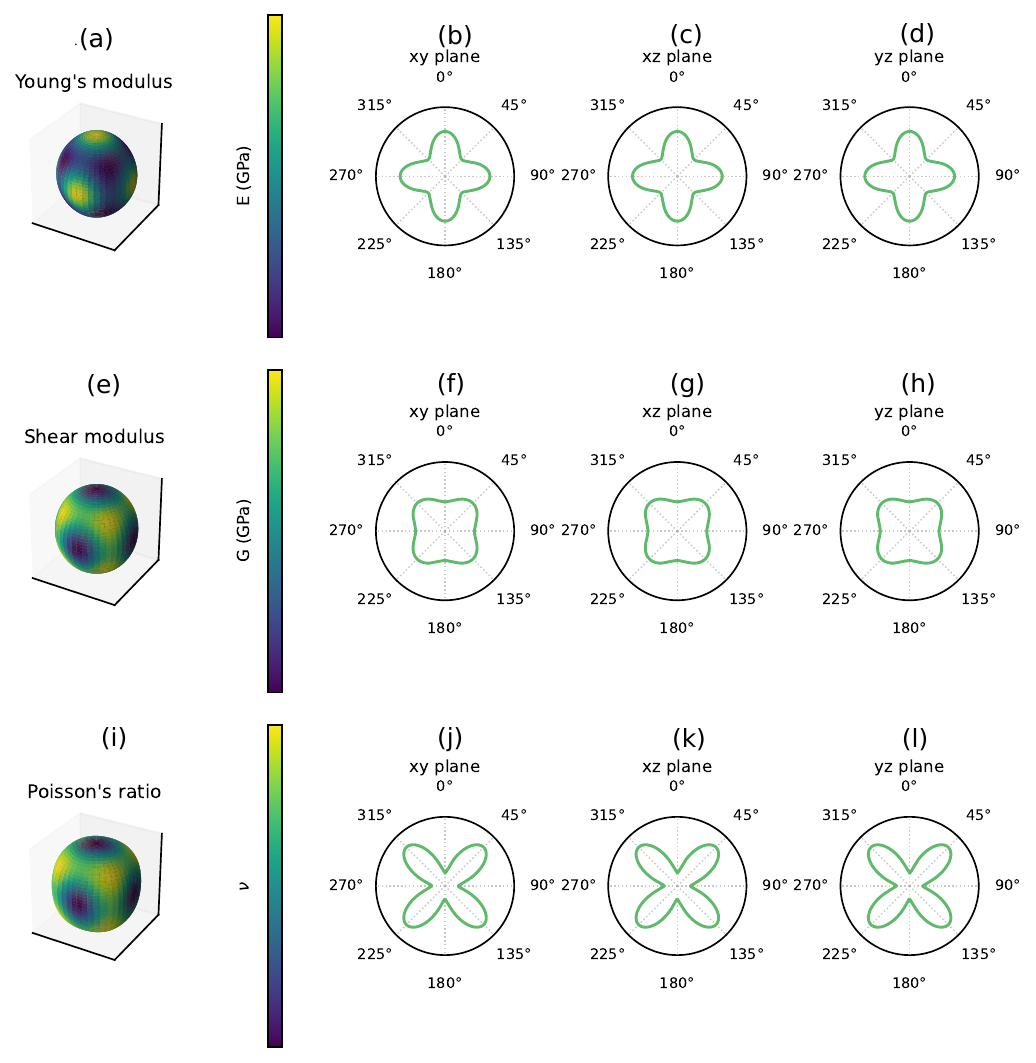}
    \caption{Orientation-dependent elastic response of cubic Mg\textsubscript{3}AsBr\textsubscript{3}.
           \textbf{(a)} 3D surface and  \textbf{(b} - \textbf{(d)} 2D polar projection of Young’s modulus
           $E(\mathbf{n})$ in XY, YZ and YZ crystallographic planes, respectively.  
           \textbf{(e)}–\textbf{(h)} Corresponding representations of the shear modulus
           $G(\mathbf{n})$.
           \textbf{(i)}–\textbf{(l)} Polar projections of the Poisson ratio
           $\nu(\mathbf{n})$ in the same planes.The colour bar encodes the magnitude of each property in GPa.}
  \label{fig:elastic_Mg3AsBr3}
\end{figure}

\begin{figure}
    \centering
    \includegraphics[width=1\linewidth]{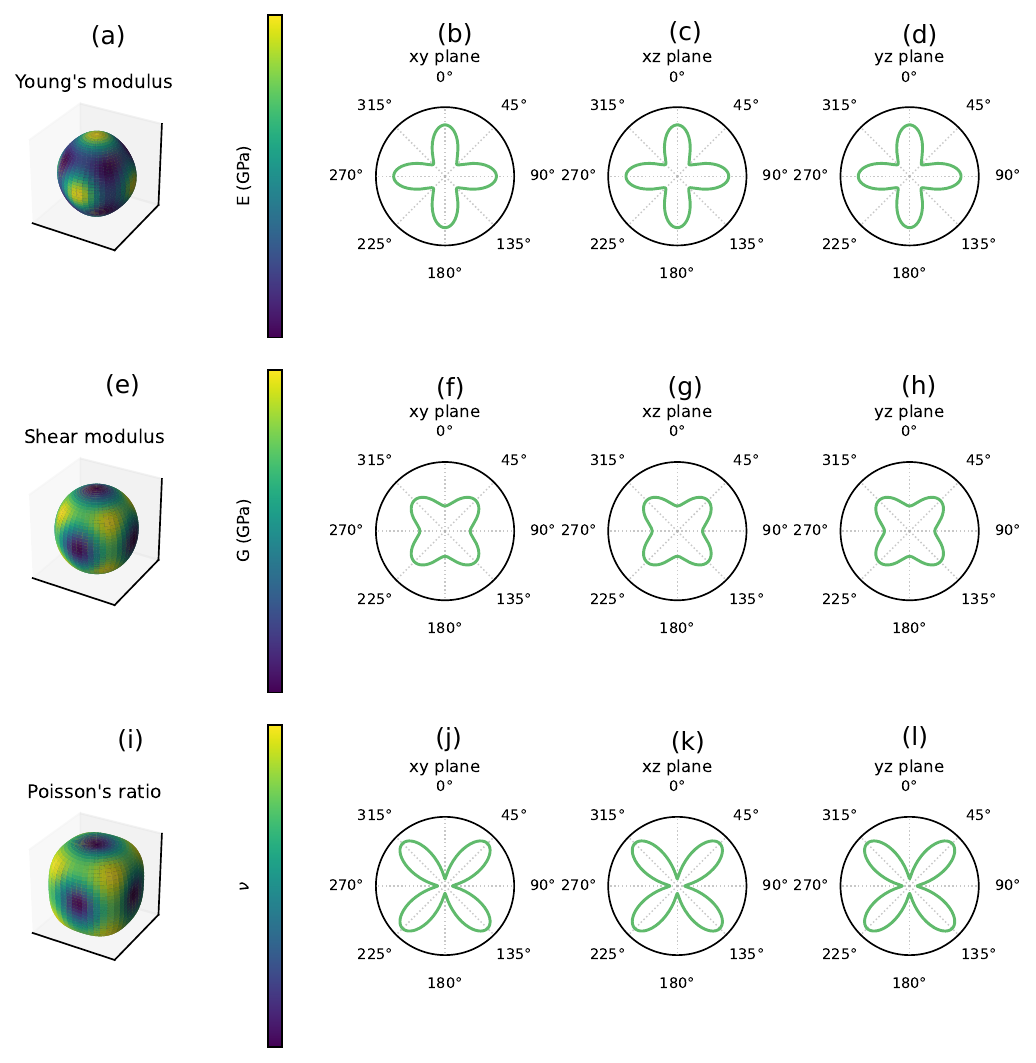}
    \caption{Orientation-dependent elastic response of cubic Mg\textsubscript{3}SbBr\textsubscript{3}.
           \textbf{(a)}–\textbf{(d)} Three-dimensional surface and two-dimensional polar projections of the Young’s modulus
           $E(\mathbf{n})$ in different crystallographic planes, respectively.
           \textbf{(e)}–\textbf{(h)} Corresponding representations of the shear modulus
           $G(\mathbf{n})$.
           \textbf{(i)}–\textbf{(l)} Polar projections of the Poisson ratio
           $\nu(\mathbf{n})$ in the same planes.The colour bar encodes the magnitude of each property in GPa.}   
           \label{fig:elastic_Mg3SbBr3}
\end{figure}

\begin{figure}
    \centering
    \includegraphics[width=1\linewidth]{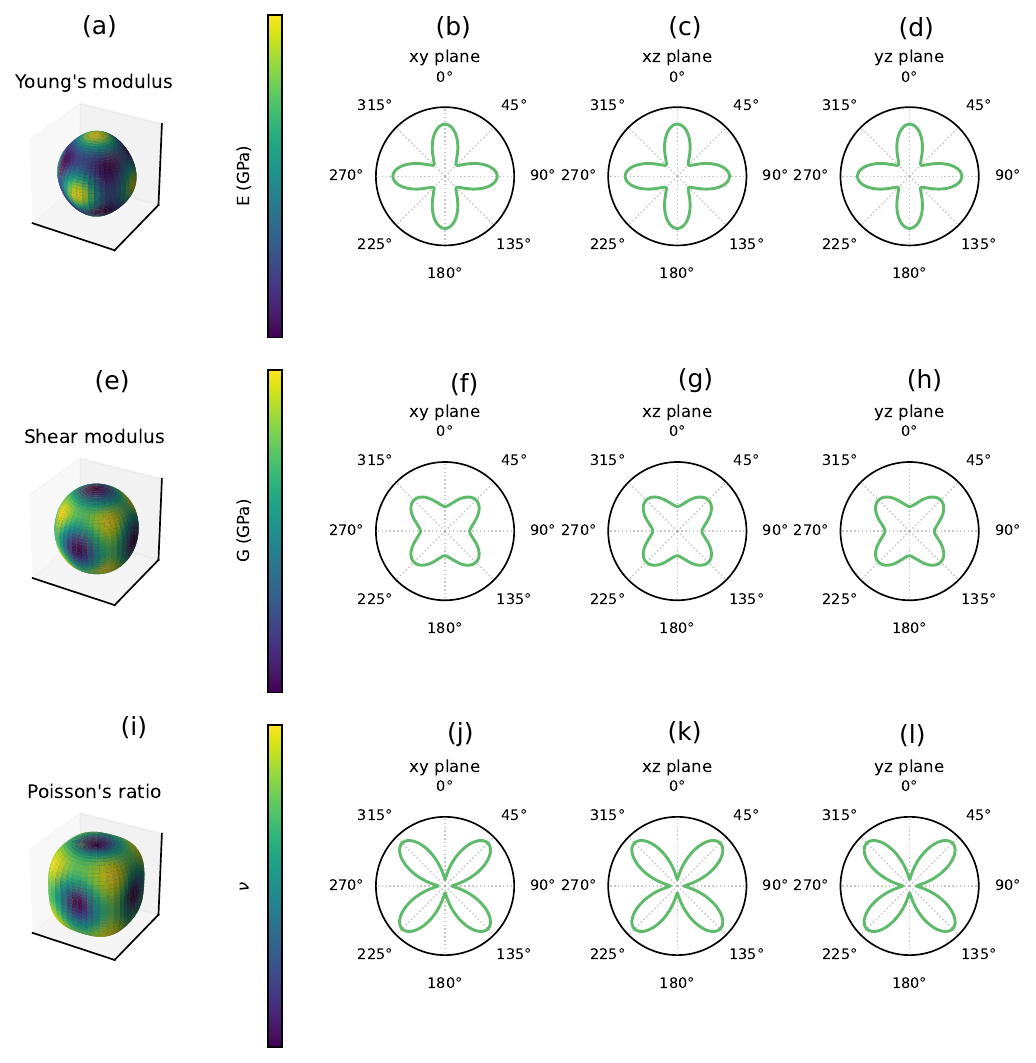}
    \caption{Orientation-dependent elastic response of cubic Mg\textsubscript{3}BiBr\textsubscript{3}.
           \textbf{(a)}–\textbf{(d)} Three-dimensional surface and two-dimensional polar projections of the Young’s modulus
           $E(\mathbf{n})$ in different crystallographic planes, respectively.
           \textbf{(e)}–\textbf{(h)} Corresponding representations of the shear modulus
           $G(\mathbf{n})$.
           \textbf{(i)}–\textbf{(l)} Polar projections of the Poisson ratio
           $\nu(\mathbf{n})$ in the same planes.The colour bar encodes the magnitude of each property in GPa.}   
           \label{fig:elastic_Mg3BiBr3}
\end{figure}

The elastic anisotropy of a material is crucial for understanding its mechanical behavior under complex stress states and has implications for various phenomena, including crack propagation, plastic deformation, and phase stability. From (a)-(l) of Figure-\ref{fig:elastic_Mg3AsBr3},\ref{fig:elastic_Mg3SbBr3},\ref{fig:elastic_Mg3BiBr3}, the anisotropies of Young's modulus, shear modulus, and Poisson's ratio in single crystals. For Mg$_3$AsBr$_3$, the anisotropy ratios are 1.072 for Young's modulus and 1.089 for shear modulus, and 1.248 for Poisson's ratio, while for Mg$_3$SbBr$_3$, these values increase to 1.166, 1.205 and  1.576, respectively. For Mg$_3$BiBr$_3$ the anisotropy ratios are 1.182 for Young's modulus and 1.225 for shear modulus, and 1.639 for Poisson's ratio. This results can be seen from the Figure-\ref{fig:elastic_Mg3AsBr3},\ref{fig:elastic_Mg3SbBr3},\ref{fig:elastic_Mg3BiBr3}. Since the anisotropies of Young's and shear modulus are close to 1 thus Figure-\ref{fig:elastic_Mg3AsBr3},\ref{fig:elastic_Mg3SbBr3},\ref{fig:elastic_Mg3BiBr3} (a and e) show spherical 3D plots, also the polar plots (b-d and f-h) further verify the symmetry. In case of Poission ratio, the anisotropy value increases with pnictogen size, which can be seen from Figure-\ref{fig:elastic_Mg3AsBr3},\ref{fig:elastic_Mg3SbBr3},\ref{fig:elastic_Mg3BiBr3} (i).The polar projection of the symmetric directional dependence of Poission's ratio can be seen for all three materials from Figure-\ref{fig:elastic_Mg3AsBr3},\ref{fig:elastic_Mg3SbBr3},\ref{fig:elastic_Mg3BiBr3} (j-l). This enhanced anisotropy with increasing pnictogen size in the compound indicates greater directional dependence of mechanical properties, which could influence behavior under complex stress states and has implications for potential applications where directional mechanical response is important. Despite their cubic symmetry, all of the compounds exhibit some degree of elastic anisotropy, as quantified by the universal elastic anisotropy index ($A^U$) of 0.01 for Mg$_3$AsBr$_3$, 0.04, and 0.050 for Mg$_3$BiBr$_3$. These values are relatively low compared to many other cubic materials, indicating near-isotropic elastic behavior, which is advantageous for applications requiring uniform mechanical response \cite{Ranganathan2008}. A more traditional measure of elastic anisotropy in cubic crystals is the Zener anisotropy ratio ($A = 2C_{44}/(C_{11}-C_{12})$), which equals 0.92 for Mg$_3$AsBr$_3$ and 0.83, Mg$_3$SbBr$_3$, and 0.82 for Mg$_3$BiBr$_3$. These values, being close to unity -- especially for the Mg$_3$AsBr$_3$  compound-- further confirm the relatively isotropic elastic nature of these materials, although the slightly lower value for Mg$_3$BiBr$_3$ suggests increasing anisotropy with larger pnictogen atoms \cite{Zener1948}. The elastic anisotropy has important implications for microscopic deformation mechanisms and potential failure modes. Materials with significant elastic anisotropy often exhibit preferential slip systems and anisotropic fracture behavior. The relatively isotropic nature of the Mg$_3$ZBr$_3$ compounds suggests more uniform mechanical response and potentially more predictable failure modes, advantageous for structural applications \cite{Chung1967}.
\begin{figure}[h!]
    \centering
    \includegraphics[width=1.0\textwidth]{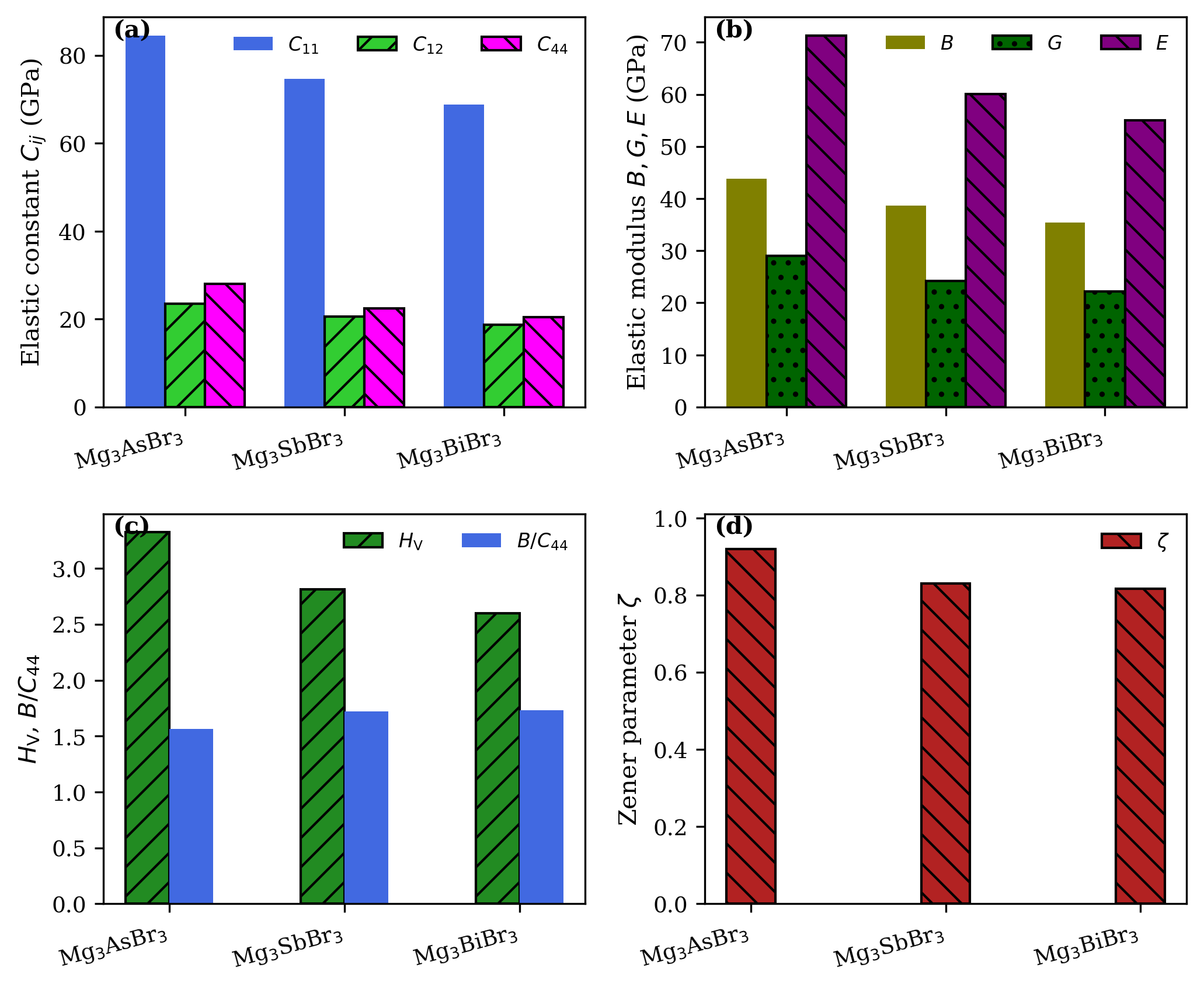}
    \caption{Calculated mechanical properties of Mg$_3$AsBr$_3$, Mg$_3$SbBr$_3$, and Mg$_3$BiBr$_3$. 
    (a) Single-crystal elastic constants $C_{ij}$ ($C_{11}$, $C_{12}$, $C_{44}$); 
    (b) bulk modulus ($B$), shear modulus ($G$), and Young’s modulus ($E$); 
    (c) Vickers hardness ($H_V$) and ratio $B/C_{44}$; and 
    (d) Zener anisotropy parameter ($\zeta$). 
   }
    \label{fig:elastic_properties_bar_plot}
\end{figure}

The decreasing elastic constrants of Figure-\ref{fig:elastic_properties_bar_plot}(a) , elastic  moduli in Figure-\ref{fig:elastic_properties_bar_plot} (b), Vickers hardness in Figure-\ref{fig:elastic_properties_bar_plot} (c) and Zener parameter in Figure-\ref{fig:elastic_properties_bar_plot} (d) can be observed. A high B/C44 ratio generally suggests better machinability, and for these materials B/C44 ratio increases with pnictogen atom size as can be observed from Figure-\ref{fig:elastic_properties_bar_plot}(c). All these trends can be described with increasing pnictogen size (As $\rightarrow$ Sb $\rightarrow$ Bi) can be attributed to the decrease of bond strength with increasing ionic radius from As$^{3+}$ (0.58 \AA) to Sb$^{3+}$ (0.76 \AA) to Bi$^{3+}$ (1.03 \AA)\cite{Pauling1960}. The valence electron configurations (As: 4s$^2$4p$^3$, Sb: 5s$^2$5p$^3$, Bi: 6s$^2$6p$^3$) also influence bonding characteristics; more diffuse orbitals of heavier pnictogens lead to weaker overlap with Br orbitals and consequently reduced bond strength and elastic moduli \cite{Harrison1989}. The increasing polarizability from As to Sb to Bi results in more easily distorted electron distributions under stress, contributing to decreased elastic moduli and increased Poisson's ratio \cite{Burdett1982}.

\section{Device Characteristics of Mg\textsubscript{3}ZBr\textsubscript{3} PIN Photodiodes} \label{sec:mg3zbr3-device}

\subsection{Structure and Material Parameters}

\begin{figure}
    \centering
    \includegraphics[width=1\linewidth]{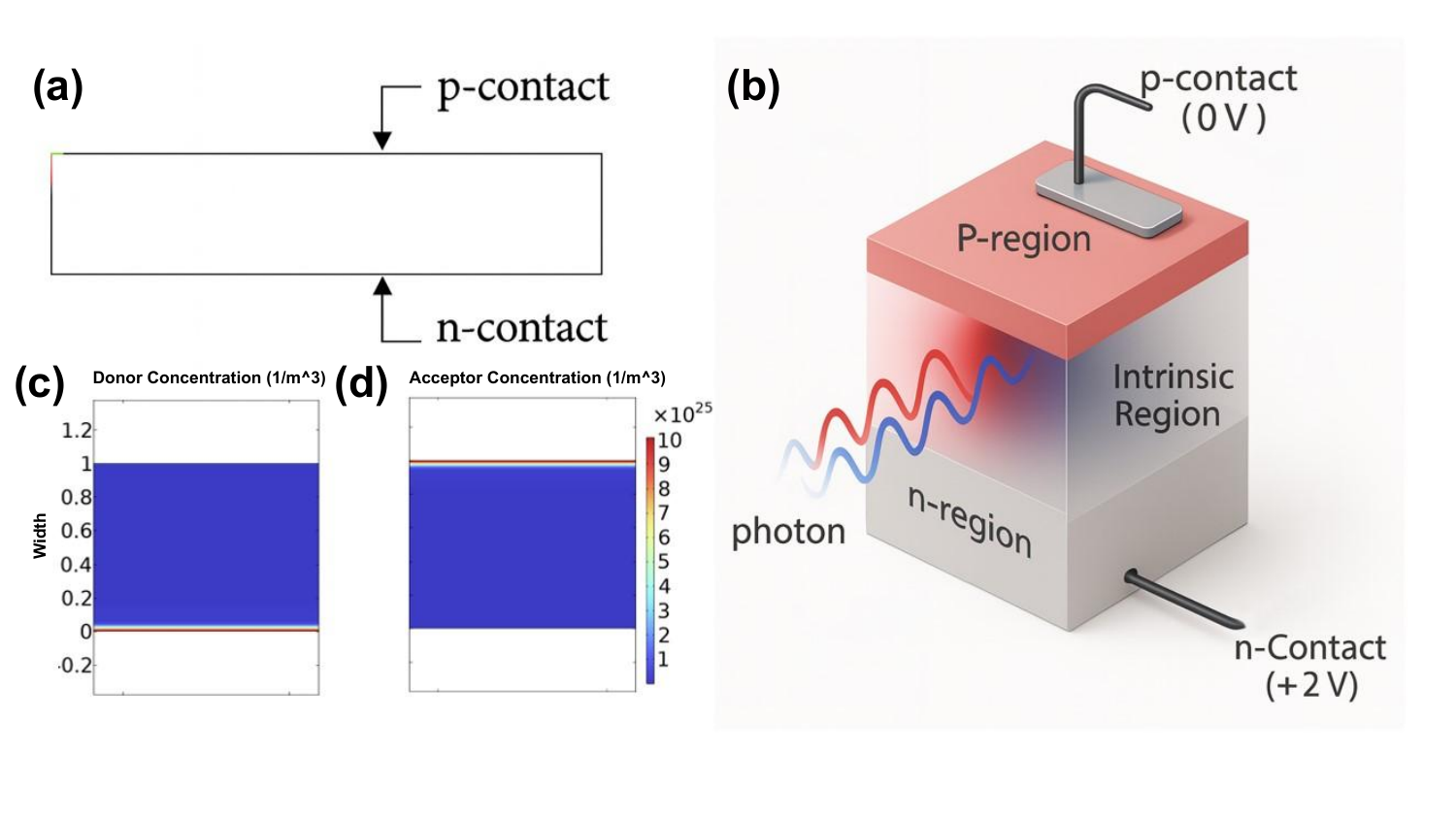}
\caption{Schematic and simulated characteristics of the Mg$_3$ZBr$_3$ (Z = As, Sb, Bi) PIN photodiode under photoconductive mode. (a)the device geometry
is a simple rectangle with a p-contact on the top surface and an n-contact on the bottom
surface.  (b) shows the 3D structural schematic of the photodiode, highlighting the P, intrinsic, and N regions along with metal contacts and applied bias. (c) presents the donor concentration. (d) displays the acceptor concentration.}
    \label{fig:photodiodescematic}
\end{figure}

Rectangular \ce{Mg3ZBr3} PIN devices, each 5\,\(\mu\)m wide and normalised to an out‑of‑plane depth of 1\,\(\mu\)m, were analysed at 293\,K using the \emph{Optoelectronics} interface of \textsc{COMSOL}~Multiphysics. The simulated stack—illustrated in \autoref{fig:photodiodescematic} (a) and (b)— consists of $p$‑ and $n$‑type cladding layers separated by an intrinsic absorber. A uniform dopant density of \(1\times10^{20}\,\mathrm{cm^{-3}}\) was applied to the contact regions \autoref{fig:photodiodescematic}(c) and (d). Transport calculations employ room‑temperature mobilities, a single mid‑gap Shockley–Read–Hall trap with \(\tau_{n}=\tau_{p}=1\)\, ns, and spontaneous radiative recombination characterised by \(\tau_{\text{spon}}=2\)\, ns. All parameters used in calculations are compiled in Table~\ref{tab:params}, which has been adapted from the DFT part of this study.  The electrostatics and transport follow the standard Poisson/continuity system \cite{SzeNg2007,VanRoosbroeck1950}:
\begin{align}
\nabla\!\cdot\!(\varepsilon\nabla\varphi) &= -q\big(p-n+N_D^+-N_A^-\big), \\
\nabla\!\cdot\!\mathbf{J}_n &= q(R-G),\quad \mathbf{J}_n = q\mu_n n\nabla\varphi + q D_n \nabla n,\\
-\nabla\!\cdot\!\mathbf{J}_p &= q(R-G),\quad \mathbf{J}_p = q\mu_p p\nabla\varphi - q D_p \nabla p,
\end{align}
with Einstein relations $D_{n,p}=\mu_{n,p}k_BT/q$. Recombination includes SRH, radiative, and Auger terms \cite{ShockleyRead1952,RoosbroeckShockley1954}:
\begin{equation}
R=R_{\mathrm{SRH}}+R_{\mathrm{rad}}+R_{\mathrm{Auger}},\quad
\begin{cases}
R_{\mathrm{SRH}}=\dfrac{np-n_i^2}{\tau_p(n+n_1)+\tau_n(p+p_1)},\\[6pt]
R_{\mathrm{rad}}=B\,(np-n_i^2),\\[4pt]
R_{\mathrm{Auger}}=(C_n n + C_p p)\,(np-n_i^2).
\end{cases}
\end{equation}

  Midgap SRH trap with $\tau_n=\tau_p=1$~ns was used (all as compiled in this work). For halide-perovskite radiative recombination we adopted $B=1.0\times10^{-10}$~cm$^3$s$^{-1}$, within the experimentally established room-temperature range $k_2\approx(0.6\text{--}14)\times10^{-10}$~cm$^3$s$^{-1}$ \cite{JohnstonHerz2016,Richter2016}. Auger coefficients were set to $C_n=C_p=1.0\times10^{-28}$~cm$^6$s$^{-1}$, consistent with reported $k_3\sim(0.2\text{--}1.6)\times10^{-28}$~cm$^6$s$^{-1}$ and first-principles/experimental determinations near $7.3\times10^{-29}$~cm$^6$s$^{-1}$ \cite{JohnstonHerz2016,Shen2018AEM}. Reverse-bias LED operation was evaluated at $V_{\mathrm{app}}=2$~V. A mapped 1D mesh (minimum element $\lesssim \min\{L_D, L_T\}/10$) was used; a zero-bias equilibrium solve initialized the carriers, followed by stationary Newton iterations at the target bias with relative tolerance $10^{-6}$. Dark current is the current that flows through the photodetector in the absence of incident light, primarily due to thermal generation of carriers. For this simulation, the effect of dark current was omitted for this study\cite{Ollearo2021_NatComm}.

\begin{table}[htbp]
\centering
\footnotesize
\caption{Material parameters adopted in the drift–diffusion simulations for the \ce{Mg3ZBr3} family.}
\label{tab:params}
\begin{tabular}{@{}lccccl@{}}
\toprule
\multirow{2}{*}{Property} & \multirow{2}{*}{Symbol} & \multirow{2}{*}{Unit} &
\multicolumn{3}{c}{\ce{Mg3ZBr3}} \\ \cmidrule(lr){4-6}
 &  &  & $Z=$As & $Z=$Sb & $Z=$Bi \\ \midrule
Relative permittivity                 & $\varepsilon_r$              & --                               & 4.57 & 5.11 & 5.7 \\
Band gap                              & $E_\text{g}$                 & eV                               & 2.06 & 1.65 & 1.52 \\
Valence–band effective DOS            & $N_\text{v}$                 & cm$^{-3}$                        & $2.78\times10^{19}$ &  $2.4\times10^{19}$ & $2.2\times10^{19}$ \\
Conduction–band effective DOS         & $N_\text{c}$                 & cm$^{-3}$                        &  $4.72\times10^{18}$ & $4.7\times10^{18}$ & $4.7\times10^{19}$ \\
Electron mobility                     & $\mu_n$                      & cm$^{2}\,\text{V}^{-1}\,\text{s}^{-1}$ & 57.86 & 30.22 & 59.2 \\
Hole mobility                         & $\mu_p$                      & cm$^{2}\,\text{V}^{-1}\,\text{s}^{-1}$ & 4.87 & 3.54 & 2.21 \\
Refractive index (real part)          & $n$                          & --                               & 2.34 & 2.14 & 2.28 \\
Refractive index (real part)          & $k$                          & --                               & --- & --- & 0.00 \\ \bottomrule
\end{tabular}
\end{table}


\begin{figure}
    \centering
    \includegraphics[width=1.0\linewidth]{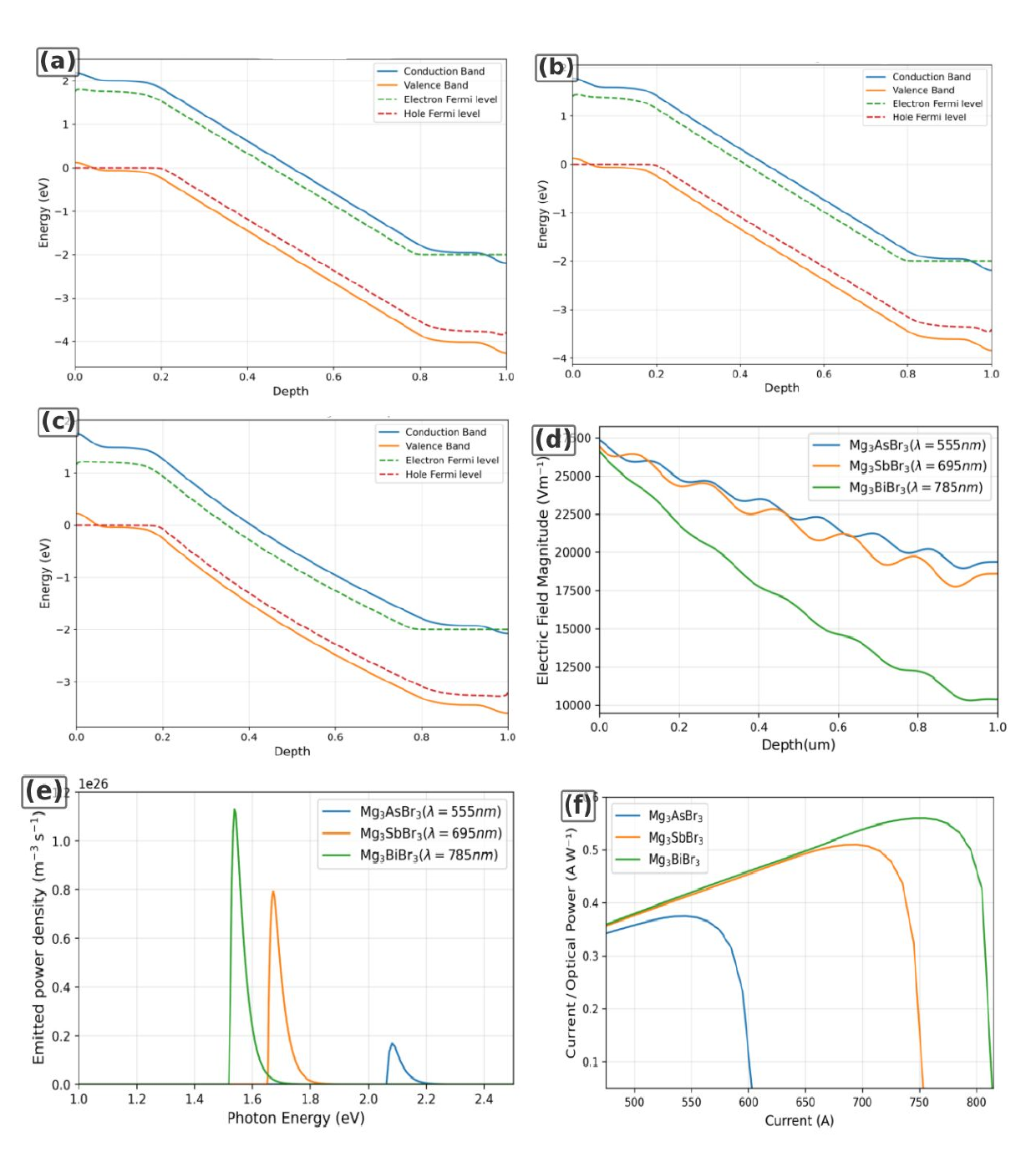}
    \caption{Simulated device characteristics of \ce{Mg3AsBr3}, \ce{Mg3SbBr3}, and \ce{Mg3BiBr3} PIN photodiodes.
    (a–c) Conduction- and valence-band edges vs.\ position.
    (d) Electric-field magnitude in the intrinsic layer at the wavelength of peak responsivity shows an approximately linear decay consistent with a uniformly depleted \(1\,\mu\text{m}\) absorber under modest external bias. (e) Spontaneous emission spectra under monochromatic illumination (\(\lambda=725\,\text{nm}\)).}

    \label{fig:photodiode}
\end{figure}

\subsection{Spontaneous Emission and Recombination}
The effectiveness of the PIN structure in photodiode devices comes from the sloping conduction and valence band profiles, which can be observed from Figure~\ref{fig:photodiode} (a) to (c), with the highest energy at the p-contact and the lowest at the n-contact.A lso from the figure, it can be observed that in the i-region of the device, the electron quasi-Fermi level is positioned beneath the conduction band, and the hole quasi-Fermi level is located above the valence band. This energy alignment ensures that the conduction band is empty and the valence band is filled, allowing efficient absorption of incoming photons. When light is absorbed and generates an electron–hole pair, the built-in electric field drives the electron toward the n-contact and the hole toward the p-contact.In addition to that, Figure~\ref{fig:photodiode} (d) presents the variation in electric field magnitude across the device for different incident wavelengths. The strength of the field decreases during propagation, showing a nearly linear trend because of the 1 μm thickness. With a thicker absorbing layer, the characteristic exponential decay of the electric field would be observed more distinctly.

Figure~\ref{fig:photodiode} (e) shows the simulated spontaneous emission spectra from the Mg$_3$ZBr$_3$ PIN diodes under above-bandgap illumination (the same conditions used for the responsivity curves). Under reverse bias, a fraction of the electron–hole pairs recombine radiatively in the intrinsic region, giving rise to photoluminescence (electroluminescence would occur under forward bias, but here the photon output is a result of optical pumping). The emission spectra confirm that each material’s photon emission onset corresponds closely to its bandgap energy. For the Bi-based diode, no emitted light is seen for photon energies below $\sim1.52 $ eV, consistent with the transparency of the material below the band edge. Once the incident photon energy exceeds $E_g$, radiative recombination can occur; Mg$_3$BiBr$_3$ shows an emission band peaking around 1.55–1.65 eV. Analogous behavior is observed for the Sb and As devices. Mg$_3$SbBr$_3$ (gap $\sim1.65$ eV) emits photons only above $\sim1.2$ eV, with a peak around 1.7–1.8 eV. Mg$_3$AsBr$_3$, having the largest gap, its emission begins at about 2.0 eV and peaks near 2.15 eV. In essence, each device emits a spectrum characteristic of its bandgap when pumped with super-bandgap light. The peaks of spontaneous emission align with the expected band-edge radiative transitions of these materials.
\noindent

The simulations substantiate that Mg$_3$ZBr$_3$‑family halide perovskites furnish a tunable platform for near‑IR to visible photodetection. By judicious compositional choice, one can engineer both the cut‑on wavelength and the peak responsivity without altering the device geometry. These quantitative insights furnish a roadmap for fabricating high‑gain, low‑dark‑current Mg‑based halide PIN photodiodes, and they lay the groundwork for subsequent experimental validation of this emerging semiconductor cohort.

\subsection{Spectral Responsivity and Photocurrent}
Figure~\ref{fig:photodiode} (f) compares the simulated spectral responsivity $R(\lambda)$ – defined as the photocurrent output per incident optical power (A/W) – of the Mg$_3$AsBr$_3$, Mg$_3$SbBr$_3$, and Mg$_3$BiBr$_3$ PIN photodiodes under steady monochromatic illumination. The incident light power is fixed (per unit area), and the wavelength is swept from 875 nm to 475 nm. As expected, the Mg$_3$AsBr$_3$ device shows a sharp long-wavelength cutoff just below 840 nm, corresponding to its $\sim$1.48eV bandgap; virtually no photocurrent is generated for $\lambda > E_g/hc$. In contrast, the Sb and Bi-based diodes remain photo-responsive at 875nm (1.417 eV), since this photon energy exceeds their smaller band gaps. For Mg$_3$BiBr$_3$, with the narrowest gap, the responsivity is in fact highest at the longest wavelength in the range (near 875nm), and $R(\lambda)$ gradually \textit{decreases} toward the blue end of the sweep. This behavior reflects the trade-off between photon absorption probability and photon flux: at longer wavelengths (just above the band edge), the material absorbs photons less strongly, but the fixed incident power corresponds to a higher photon flux. At shorter wavelengths, absorption in the 1~$\mu$m i-layer becomes nearly complete, but fewer incident photons (due to higher energy per photon) limit the photocurrent. It can also be observed that the peak shifted to longer wavelengths as $E_g$ decreases from As to Bi. Specifically, Mg$_3$AsBr$_3$ reaches a peak $R\approx 0.09$ A/W around $\lambda\sim720$–730 nm. Mg$_3$SbBr$_3$ shows a broader responsivity plateau peaking near $\sim800$ nm with a slightly higher maximum $R$ (on the order of $0.1$A/W). Mg$_3$BiBr$_3$, which can absorb effectively throughout the entire 875–475nm range, exhibits a relatively flat or gently sloping response that tends to peak toward the red end; for instance, $R$,$\approx0.08$–$0.09$ A/W at 875 nm and gradually diminishes to $\sim0.05$–$0.06$ A/W by 500 nm.
We note these values are lower than those of optimized or thicker photodiodes; e.g., single-crystal or tandem perovskite photodetectors often report $R\sim0.3$–$0.6$ A/W in the visible\cite{Ollearo2021_NatComm}. The modest $R$ here is attributable to the simple device design (only 1μm absorption path, no reflective or optical concentrator layers) and the assumption of unity carrier collection efficiency without internal gain. Nonetheless, the extension of spectral sensitivity into the infrared for Sb and especially Bi-based Mg$_3$ZBr$_3$ is a clear advantage. 

\section{Conclusion}

First-principles calculations identify cubic perovskite Mg$_3$ZBr$_3$ (Z = As, Sb, Bi; $Pm\bar{3}m$) as a chemically simple, lead-free halide family with predictable trends across structure, lattice dynamics, and optoelectronic response. After relaxation, the lattice parameter and equilibrium volume increase from As $\rightarrow$ Sb $\rightarrow$ Bi, consistent with the larger pnictogen size. Phonon dispersions contain no imaginary branches at 0~K, consistent with a dynamically stable cubic phase, while the spectrum softens down the group. Hybrid functional band structures yield indirect gaps of $\sim$2.06~eV (As), $\sim$1.65~eV (Sb), and $\sim$1.52~eV (Bi), with lighter electrons than holes along the fundamental transition, suggesting electron-favored transport.

Frequency-dependent optical functions follow the electronic trends. The static dielectric constant rises from $\sim$4.8 (As) to $\sim$5.9 (Bi), and the near-IR refractive index lies near 2.2--2.5. Absorption onsets track the gaps, with weak near-edge absorption typical of indirect semiconductors and rapid growth to $10^{5}$--$10^{6}$~cm$^{-1}$ at higher photon energies; visible-range absorption is appreciable for Sb and Bi. Reflectivity near the band edge is $\sim$0.15--0.19. These properties place Mg$_3$SbBr$_3$ and Mg$_3$BiBr$_3$ in the useful 1.5--1.7~eV window for thin-film optoelectronics, while Mg$_3$AsBr$_3$ targets the violet/UV.

Elastic tensors satisfy the Born criteria for all three compositions. Hill-averaged moduli decrease smoothly down the series (bulk modulus $\approx$ 44 $\rightarrow$ 35~GPa; Young's modulus $\approx$ 71 $\rightarrow$ 55~GPa), accompanied by lower Debye temperatures. Quasi-harmonic mode Gr\"uneisen parameters are large ($\bar{\gamma} \gtrsim 2.7$--3.3), consistent with strong anharmonicity and low predicted lattice thermal conductivity.

Device-level drift--diffusion simulations of $p$--$i$--$n$ stacks reproduce band-edge-limited spectral cut-ons and rank-order the responsivity with composition, confirming that this series can cover near-IR to visible wavelengths without changing the device geometry. Together with the high-$Z$ content of the Bi member, these results point to applications in photodiodes and radiation detection, and they motivate measurement of thermal transport and finite-temperature phase behavior to test the predicted anharmonicity and stability margins.

\section{Declaration of Competing Interest}
The authors declare that they have no known competing financial interests or personal relationships that could have appeared to influence the work reported in this paper.

\section{Data Availability}
The data supporting the findings of this study are available upon request from the corresponding author. 
\section{Author Contributions}
Md. Mohiuddin and Mehedi Hasan designed the study and performed the calculations, analyzed the data, and wrote the manuscript. A. Kabir administered the project and supervised the investigations by guiding the computations, analyzing the data. All authors reviewed the manuscript.

\section{Acknowledgments}
The authors acknowledge the University of Dhaka for facilitating the research environment in the Department of Physics, and Bangladesh Research and Education Network BdREN (bdren.net.bd) for the computational lab facilities.Text drafting and language polishing were assisted by a large language model (ChatGPT, OpenAI; gpt 4o model). The tool was used only for prose drafting/editing; it did not generate or analyze data, and it did not influence methodological or interpretive decisions. All authors verified the text and are responsible for all claims and conclusions.



\end{document}